\documentclass[11pt,a4paper]{article}

\usepackage{jheppub}
\usepackage{amsthm}

\usepackage{amsmath,amssymb,amsfonts,mathtools}
\usepackage{mathrsfs}
\usepackage{bm}
\usepackage{bbm}
\usepackage{physics}
\usepackage{slashed}
\usepackage{graphicx}
\usepackage{hyperref}
\usepackage{xcolor}
\usepackage{float}
\usepackage[section]{placeins}

\newcommand{\cN}{\mathcal N}

\newcommand{\cP}{\mathcal P}
\newcommand{\cQ}{\mathcal Q}

\newcommand{\eps}{\epsilon}

\newcommand{\fZ}{\mathfrak Z}
\newcommand{\fS}{\Sigma}
\newcommand{\bk}{\beta_\kappa}

\title{The Two Lives of a Massive Charged Spin-$\tfrac32$ Particle:\\
from Superstrings EFT to Supergravity}

\author[a]{Karim Benakli}
\affiliation[a]{Sorbonne Universit\'e, CNRS, Laboratoire de Physique Th\'eorique et Hautes \'Energies, LPTHE, F-75005 Paris, France}
\emailAdd{kbenakli@lpthe.jussieu.fr}

\abstract{
A charged massive spin-$\tfrac32$ field in a constant electromagnetic background admits two familiar realizations. In supergravity, where it is identified with a gravitino, its mass and charge are tied in Planck units. In the decoupled-gravity regime, the two-derivative Fierz--Pauli system describing the first massive open-string modes instead allows arbitrary mass and charge. The gyromagnetic ratio $g=2$ is achieved in both cases through non-minimal Pauli couplings, but these take different forms: they are symmetric in the two chiral sectors in the supergravity system, and chiral-asymmetric in the flat-space one. We construct a continuous family of Fierz--Pauli systems interpolating between these two endpoints.

We show that the corresponding second-order equations always reproduce $g=2$ at linear order, but generically contain a chiral $F^2$ term at quadratic order. Its coefficient vanishes precisely at the two endpoint theories. Requiring compatibility with dynamical gravity, or closure for non-constant electromagnetic field strength, then selects the supergravity endpoint. The result clarifies the domains of applicability of the two systems and shows that the maximally asymmetric organization of the decoupled theory cannot be extended unchanged once gravity is dynamical. For comparison, we also analyze a charged massive spin-$2$ field on Einstein--Maxwell backgrounds and exhibit the corresponding closure of the associated lower-spin chain.
}
\begin{document}
\maketitle
\flushbottom

\section{Introduction}
\label{sec:intro}

A massive spin-$\tfrac32$ particle has \(2s+1=4\) physical degrees of freedom, but the vector--spinor field
\(\Psi_\mu\) used to describe it contains more components than that. Its equations of motion must therefore come
with constraints that remove the unphysical lower-spin sector. The equations of motion together with this
constraint chain form a Fierz--Pauli system~\cite{Fierz:1939ix,Rarita:1941mf}.

For a charged massive spin-$\tfrac32$ field in a constant electromagnetic background, two consistent Fierz--Pauli
organizations are known. One belongs to a decoupled regime\footnote{In the following, ``decoupled'' always means
decoupled from gravity.}, where gravity is neglected~\cite{Benakli:2021jxs,Benakli:2022edf,Benakli:2023aes}. The
other is the reduced \(\cN=2\) Einstein--Maxwell supergravity system, where the same type of massive charged
spin-$\tfrac32$ degree of freedom is coupled to both electromagnetism and dynamical gravity~\cite{Das:1976ct}. Both
are consistent in the regime in which they are derived, but they are organized in markedly different ways. Both also
have the expected gyromagnetic ratio \(g=2\)~\cite{Ferrara:1992yc}. This naturally raises the following question:
\begin{quote}
If we start from a decoupled Fierz--Pauli system and couple it to dynamical gravity,
do we recover the supergravity organization? Conversely, in the decoupling limit \(M_P\to\infty\), do we recover the
homogeneous/sourced system?
\end{quote}

What makes this question nontrivial is that the two Fierz--Pauli systems are organized in visibly different ways.
In \(\cN=2\) supersymmetry~\cite{Fayet:1975yi}, the spin-$2$ graviton and the spin-$\tfrac32$ gravitino belong to
the same multiplet. Since only the gravitino can be charged, the corresponding \(U(1)\) does not commute with
supersymmetry: it is an \(R\)-symmetry. In \(\cN=2\) supergravity~\cite{Freedman:1976xh,Deser:1977uq,Ferrara:1976fu},
gauging this \(R\)-symmetry is highly constrained and leads to a unique form of the charged gravitino equations~\cite{Das:1976ct}.
These equations are symmetric between the two chiral sectors: the Pauli magnetic coupling appears in both, and is
organized through the self-dual electromagnetic tensor. Moreover, the system requires a tuning between mass, charge,
and curvature~\cite{Das:1976ct,Deser:1977uq}. This tuned locus was later shown to be required by causal propagation on
Einstein--Maxwell backgrounds~\cite{Deser:2001dt}. For unit electric charge, however, it corresponds to a Planckian
mass, which makes this description of limited direct use for phenomenological spin-$\tfrac32$ applications.

The search for equations describing the propagation of spin-$\tfrac32$ particles in electromagnetic backgrounds goes
back at least to Dirac~\cite{Dirac:1936tg}. These developments revealed two aspects of the same underlying problem:
the construction of a causal Fierz--Pauli system. The first is to find a set of constraints that is compatible both
internally and with the equations of motion. The second is to ensure that these constraints do not degenerate for
special values of the electromagnetic background. The failure of minimal coupling to satisfy these requirements is the
Velo--Zwanziger problem: the essential question is whether the reduced system on the physical subspace remains
consistent and hyperbolic~\cite{Johnson:1960vt,Velo:1969bt,Velo:1969txo}.

To address the first aspect, Fierz and Pauli advocated deriving both the equations of motion and the constraints from
a variational principle. In practice, the search for satisfactory Lagrangians involving only a single massive
spin-$\tfrac32$ field has led to different realizations of the Fierz--Pauli equations: an implicit approach from
which one can extract an order-by-order expansion in \(e|F|/m^2\)~\cite{Porrati:2009bs}, and a formulation employing
auxiliary fields whose elimination generates a non-polynomial dependence on \(F_{\mu\nu}\)~\cite{Delplanque:2024xst}.
By contrast, if one allows additional dynamical fields---which is less satisfactory for the original single-field
purpose---open string field theory naturally leads to the decoupled constant-field system studied in
\cite{Benakli:2021jxs,Benakli:2023aes,Benakli:2022edf}. The second aspect of the Velo--Zwanziger problem is avoided
there by maintaining the Rarita--Schwinger form of the primary constraint, namely the vanishing gamma-trace of the
vector--spinor. We shall impose that same primary constraint throughout this work. Within the constant-field,
two-derivative framework considered here, the resulting decoupled Fierz--Pauli system was shown to be unique at
\(g=2\)~\cite{Benakli:2024aes}. Departures from \(g=2\), of possible phenomenological interest, require an effective
field-theory extension~\cite{Benakli:2025xao}.

In the decoupled regime, the consistent constant-field Fierz--Pauli system derived from the open-string-inspired
construction has the distinctive feature that the non-minimal Pauli coupling appears only in one of the two
spin-$\tfrac32$ chiral equations of motion and in the corresponding secondary constraint. Mass and charge are then
independent parameters. This is in clear contrast with the reduced \(\cN=2\) supergravity system, where the Pauli
coupling is symmetric between the two chiralities and the parameters are tied by the supergravity relation.

The purpose of the present work is to compare these two organizations directly at the level of the covariant
Fierz--Pauli chain. Rather than treating the flat-space and supergravity systems as unrelated, we introduce a reduced
two-component interpolating family parameterized by \(\beta_\kappa\), whose endpoints reproduce the decoupled and
reduced \(\cN=2\) forms. This family keeps the same field content and primary constraints while reorganizing the
Pauli sector. The central questions are then the following: starting 
from the decoupled system at \(\beta_\kappa=0\),
does the lower-spin chain still close once curvature is included, and why is the mass--charge relation
relaxed when gravity is decoupled?

Our first result is that on Einstein--Maxwell backgrounds, the Fierz--Pauli system with generic \(\beta_\kappa\)
develops a dressed algebraic obstruction to closure at the next step of the constraint chain. The divergence of the
reduced equations produces an Einstein-tensor channel from curvature commutators and, simultaneously, a quadratic
electromagnetic channel from the Pauli sector. On Einstein--Maxwell backgrounds the former becomes proportional to the
Maxwell stress tensor. Closure requires the Pauli sector to match the same rank-two tensor structure.
This matching yields the tuned relation
\begin{equation}
M^2=\frac{2\beta_\kappa e^2}{\kappa^2},
\end{equation}
and at the symmetric endpoint \(\beta_\kappa=1\) reproduces the standard \(\cN=2\) supergravity locus.

Our second result is that the same mechanism becomes completely explicit at the level of the exact second-order
equations on Einstein--Maxwell backgrounds with covariantly constant field strength. The tuned locus cancels the
tensorial \(T^{(F)}_{\mu\nu}\) channel exactly. The linear Zeeman couplings remain universal throughout the family and
always give \(g=2\). The distinction between the interpolating theories appears only at quadratic order: for all real
\(\beta_\kappa\) except the two endpoints \(\beta_\kappa=0\) and \(\beta_\kappa=1\), the tuned second-order equations
retain a chiral scalar \(F^2\) term proportional to \(\beta_\kappa(1-\beta_\kappa)\), whereas this term vanishes
identically at the endpoints.

Our third result is that allowing non-constant electromagnetic backgrounds provides a stronger test. The secondary
divergences remain unchanged, but the next step in the chain acquires explicit \(\nabla F\) terms. For generic
\(\beta_\kappa\), these produce a naked lower-spin obstruction which is not absorbed by the dressed Einstein--Maxwell
structure. At the symmetric endpoint \(\beta_\kappa=1\), however, the naked \(\partial F\) term cancels exactly,
leaving only the physical Maxwell current. Thus the non-constant-field analysis does not merely preserve the tuned
locus: it singles out the symmetric supergravity endpoint within the interpolating family.

A useful way to interpret these results is to regard the flat-space homogeneous/sourced form as the truncation of a
more complicated system involving additional fields and/or higher-order operators in a derivative or mass-dimension
expansion. In the decoupled constant-field regime this organization is consistent. Once gravity becomes
dynamical, however, the divergence algebra probes tensor channels that are sensitive to the chiral structure of the
Pauli coupling. If one insists on a two-derivative effective description while allowing both the electromagnetic and
gravitational backgrounds to participate dynamically, the symmetric completion is forced, because only it can match
the Einstein--Maxwell stress-tensor channel and the electromagnetic divergence couplings generated by curvature.

To clarify what is special about spin-\(\tfrac32\), we also analyze the bosonic spin-2 comparison. For neutral
spin-2 on curved backgrounds, naive covariantization fails because the divergence develops curvature obstructions; on
Einstein spaces, the derivative-of-Ricci terms disappear, but a residual Weyl-coupled obstruction remains. Within the
standard local algebraic class, this Weyl term can be canceled, and the subsidiary system then closes without any
mass-curvature tuning. In particular, there is no residual obstruction on conformally flat Einstein backgrounds.

The charged spin-2 equations of motion were derived in
\cite{Argyres:1989cu,Porrati:2010hm,Benakli:2021jxs,Benakli:2022ofz,Benakli:2022edf}. In contrast with the
spin-\(\tfrac32\) case, they are a relatively direct charged extension of the massive spin-2 Fierz--Pauli system,
supplemented by the non-minimal Pauli term required to obtain gyromagnetic ratio \(g=2\), much as for massive
vectors. We show that on Einstein backgrounds with covariantly constant electromagnetic field strength, the charged
spin-2 subsidiary chain can be made to close exactly. However, the mechanism is qualitatively different from the
spin-\(\tfrac32\) one: the electromagnetic sector adds its own derivative obstruction, which can be removed, but the
divergence never develops an algebraic Einstein--Maxwell stress-tensor channel. This difference can be traced to the
order of the kinetic operator: in the spin-\(\tfrac32\) case the first-order system allows the constraint chain to
close on an algebraic rank-two tensor, whereas for spin-2 the second-order structure leaves the obstruction
intrinsically differential. This explains why a tuned Einstein--Maxwell locus arises for spin-\(\tfrac32\) but not
for spin-2.

The paper is organized as follows. Section~\ref{sec:beta_kappa_reduced_system} introduces the decoupled and reduced
\(\cN=2\) Fierz--Pauli chains and the interpolating \(\beta_\kappa\)-family.
Section~\ref{sec:beta_kappa_closure_core} analyzes the closure of the reduced chain, derives the dressed
Einstein--Maxwell obstruction, and obtains the tuned locus; it also discusses the neutral limit and its relation to
Buchdahl's condition. Section~\ref{sec:constant_field_second_order} derives the exact constant-field second-order
equations on Einstein--Maxwell backgrounds, shows the universality of \(g=2\), and isolates the residual chiral
scalar \(F^2\) term together with its rigidity under simple first-order counterterms. Section~\ref{sec:nonconstant_F}
studies non-constant electromagnetic backgrounds and shows that the \(\partial F\) test selects the symmetric
endpoint \(\beta_\kappa=1\). Section~\ref{sec:spin2_gravity_obstruction} recasts the neutral spin-2 comparison on
curved backgrounds in the subsidiary-chain language used throughout the paper, and
section~\ref{sec:spin2_charged_curved} extends that comparison to the charged case. We conclude in
section~\ref{sec:conclusions}. Many useful technical conventions and identities are collected in the appendices.

\section{Decoupled, reduced \texorpdfstring{$\cN=2$}{N=2}, and interpolating Fierz--Pauli chains}
\label{sec:beta_kappa_reduced_system}

We work in four dimensions with mostly-minus signature and use two-component Weyl notation. The vector--spinor is decomposed as
\begin{equation}
\Psi_m=
\begin{pmatrix}
\chi_{m\alpha}\\[1mm]
\bar\lambda_m{}^{\dot\alpha}
\end{pmatrix},
\qquad
\gamma^m=
\begin{pmatrix}
0 & \sigma^m_{\alpha\dot\alpha}\\
\bar\sigma^{m\,\dot\alpha\alpha} & 0
\end{pmatrix}.
\end{equation}
The gauge-covariant derivative is
\begin{equation}
D_m \equiv \partial_m + i e A_m,
\end{equation}
and throughout this section the electromagnetic field strength \(F_{mn}\) is taken to be constant.

Our goal is to compare two realizations of the Fierz--Pauli constraint chain for a massive charged spin-\(\tfrac32\) field:
\begin{itemize}
\item the \emph{decoupled} system, in which gravity does not enter the constraint algebra and the dynamics is specified directly by equations of motion together with postulated constraints,
\item the \emph{reduced \(\cN=2\) supergravity} system, in which both equations and constraints descend from a covariant Lagrangian and are rewritten in a first-order form adapted to the spin-\(\tfrac32\) sector.
\end{itemize}
The two systems differ only in the organization of their Pauli couplings and, as a result, in the structure of the constraint chain.

In both cases we impose the algebraic spin-\(\tfrac32\) conditions
\begin{equation}
\bar\sigma^m\chi_m=0,
\qquad
\sigma^m\bar\lambda_m=0,
\label{eq:primary_sigma_trace_constraints_clean}
\end{equation}
which remove the spin-\(\tfrac12\) components. Secondary constraints are obtained by acting with covariant derivatives on the equations of motion and using \eqref{eq:primary_sigma_trace_constraints_clean}. Consistency requires that this procedure close without generating additional independent lower-spin conditions.

We also introduce the self-dual and anti-self-dual combinations
\begin{equation}
(F_\pm)_{mn}\equiv F_{mn}\pm i\,\widetilde F_{mn},
\qquad
\widetilde F_{mn}\equiv \frac12\,\eps_{mnrs}F^{rs}.
\label{eq:Fpm_def_main_clean}
\end{equation}

\subsection{Decoupled versus reduced \texorpdfstring{$\cN=2$}{N=2} chains}
\label{sec:decoupled_vs_SUGRA_educed_system}

The decoupled system in flat spacetime with a constant electromagnetic background is
\begin{align}
i\,\bar\sigma^n D_n \chi_m + M\,\bar\lambda_m &= 0,
\label{eq:BBDKL_RS1_clean}
\\[1mm]
i\,\sigma^n D_n \bar\lambda_m + M\,\chi_m
&=
-\,i\,\frac{2 e}{M}\,F_{mn}\chi^n.
\label{eq:BBDKL_RS2_clean}
\end{align}
This system has an asymmetric structure: the first equation is homogeneous, while the second contains a Pauli source term. We will therefore refer to it as the \emph{homogeneous/sourced} organization.

The reduced \(\cN=2\) supergravity system instead takes a chiral-symmetric form,
\begin{align}
i\,\bar\sigma^n D_n \chi_m + M\,\bar\lambda_m
&=
-\,i\,\frac{e}{M}\,(F_+)_{mn}\,\bar\lambda^n,
\label{eq:N2_reduced_1_clean}
\\[1mm]
i\,\sigma^n D_n \bar\lambda_m + M\,\chi_m
&=
-\,i\,\frac{e}{M}\,(F_-)_{mn}\,\chi^n.
\label{eq:N2_reduced_2_clean}
\end{align}
Here both chiralities are sourced symmetrically through \(F_\pm\).
This form follows from the supergravity equations after projection onto the spin-\(\tfrac32\) sector and use of the primary constraints \eqref{eq:primary_sigma_trace_constraints_clean}.

Both systems propagate the same physical degrees of freedom and satisfy the same primary constraints. Their difference lies entirely in the organization of the Pauli couplings and, consequently, in the structure of the resulting constraint chain.

\subsection{Interpolating family}
\label{sec:interpolation_educed_system}

We introduce a one-parameter family of reduced Fierz--Pauli chains,
\begin{align}
i\,\bar\sigma^n D_n \chi_m + M\,\bar\lambda_m
&=
-\,i\,\bk\,\frac{e}{M}\,(F_+)_{mn}\,\bar\lambda^n,
\label{eq:beta_family_red_1_clean}
\\[1mm]
i\,\sigma^n D_n \bar\lambda_m + M\,\chi_m
&=
-\,i\,\frac{e}{M}
\Bigl[
2(1-\bk)\,F_{mn}
+\bk\,(F_-)_{mn}
\Bigr]\chi^n,
\label{eq:beta_family_red_2_clean}
\end{align}
where \(\bk\in\mathbb{R}\).

The endpoints\footnote{The parameter \(\bk\) is real and the family is well-defined for all \(\bk\in\mathbb{R}\). 
In practice we will often focus on the interval \(0\le \bk \le 1\), which interpolates between 
the decoupled system (\(\bk=0\)) and the reduced \(\cN=2\) system (\(\bk=1\)).}
reproduce the two systems:
\begin{equation}
\bk=0 \quad\Rightarrow\quad \text{decoupled system}, 
\qquad
\bk=1 \quad\Rightarrow\quad \text{reduced \(\cN=2\) system}.
\end{equation}

The interpolating family preserves the same field content and the same primary constraints. It changes only the organization of the Pauli sector. The question is whether this reorganization remains consistent once the full lower-spin chain is analyzed.

The secondary divergence relations follow from taking the \(\sigma\)-traces of
\eqref{eq:beta_family_red_1_clean}--\eqref{eq:beta_family_red_2_clean}. The derivation uses only the primary constraints \eqref{eq:primary_sigma_trace_constraints_clean}, the Weyl identities
\begin{equation}
\bar\sigma^{m}\sigma^n
=
2\,g^{mn}-\bar\sigma^{n}\sigma^m,
\qquad
\sigma^m\bar\sigma^{n}
=
2\,g^{mn}-\sigma^{n}\bar\sigma^{m},
\label{eq:secondary_div_sigma_identities}
\end{equation}
and the chiral trace properties
\begin{equation}
\bar\sigma^m(F_-)_{mn}\chi^n=0
\quad\text{if}\quad
\bar\sigma^m\chi_m=0,
\label{eq:secondary_div_Fminus_trace}
\end{equation}
\begin{equation}
\sigma^m(F_+)_{mn}\bar\lambda^n=0
\quad\text{if}\quad
\sigma^m\bar\lambda_m=0.
\label{eq:secondary_div_Fplus_trace}
\end{equation}

Contracting \eqref{eq:beta_family_red_2_clean} with \(\bar\sigma^m\) gives
\begin{equation}
D\!\cdot\!\bar\lambda
=
-\frac{e}{M}(1-\bk)\,\bar\sigma^m F_{mn}\chi^n,
\label{eq:secondary_div_lambdabar_final_tight}
\end{equation}
while contracting \eqref{eq:beta_family_red_1_clean} with \(\sigma^m\) yields
\begin{equation}
D\!\cdot\!\chi=0.
\label{eq:secondary_div_chi_final_tight}
\end{equation}

We therefore obtain the secondary constraints for the full family,
\begin{equation}
D\!\cdot\!\chi=0,
\qquad
D\!\cdot\!\bar\lambda
=
-\frac{e}{M}(1-\bk)\,\bar\sigma^m F_{mn}\chi^n.
\label{eq:secondary_divergences_beta_family_tight}
\end{equation}
They interpolate smoothly between the two limiting cases: for \(\bk=0\) the divergence of \(\bar\lambda_m\) is sourced, while for \(\bk=1\) both divergences are homogeneous.

It is convenient to define
\begin{equation}
\fS^{(L)}_\alpha \equiv D^m\chi_{m\alpha},
\qquad
\fS^{(R)}_{\dot\alpha}
\equiv
D^m\bar\lambda_{m\dot\alpha}
+\frac{e}{M}(1-\bk)\,\bar\sigma^m_{\dot\alpha\alpha}F_{mn}\chi^{n\alpha},
\label{eq:beta_core_Sigma_defs_tight}
\end{equation}
so that the secondary constraints take the compact form
\begin{equation}
\fS^{(L)}_\alpha=0,
\qquad
\fS^{(R)}_{\dot\alpha}=0.
\label{eq:beta_core_Sigma_zero_tight}
\end{equation}

%
%
%
%
%
%
%

%
%
%
%
%
%
%

\section{Closure of the reduced \texorpdfstring{$\beta_\kappa$}{beta-kappa} family}
\label{sec:beta_kappa_closure_core}

We now analyze the next step in the lower-spin constraint chain for the reduced interpolating system
\eqref{eq:beta_family_red_1_clean}--\eqref{eq:beta_family_red_2_clean}. 
The primary constraints \eqref{eq:primary_sigma_trace_constraints_clean} and the secondary relations
\eqref{eq:beta_core_Sigma_zero_tight} are assumed throughout. For convenience, we recall that
\begin{equation}
\fS^{(L)}_\alpha = D^m\chi_{m\alpha},
\qquad
\fS^{(R)}_{\dot\alpha}
=
D^m\bar\lambda_{m\dot\alpha}
+\frac{e}{M}(1-\bk)\,\bar\sigma^m_{\dot\alpha\alpha}F_{mn}\chi^{n\alpha}.
\label{eq:beta_core_Sigma_defs_here_clean}
\end{equation}

\subsection{Dressed secondary obstruction}

Taking the covariant divergence of \eqref{eq:beta_family_red_1_clean} and using \(D_m\bar\sigma^n=0\),
we obtain
\begin{equation}
i\,\bar\sigma^n D_n \fS^{(L)}
+i\,\bar\sigma^n[D^m,D_n]\chi_m
+M\,D\!\cdot\!\bar\lambda
=
-\,i\,\bk\,\frac{e}{M}D^m\!\left((F_+)_{mn}\bar\lambda^n\right).
\label{eq:beta_left_divergence_start_clean}
\end{equation}
Using the definition of \(\fS^{(R)}\),
\begin{equation}
D\!\cdot\!\bar\lambda
=
\fS^{(R)}
-\frac{e}{M}(1-\bk)\,\bar\sigma^m F_{mn}\chi^n,
\label{eq:beta_Ddotlambda_via_SigmaR_clean}
\end{equation}
this becomes
\begin{align}
i\,\bar\sigma^n D_n \fS^{(L)}
+M\,\fS^{(R)}
+i\,\bar\sigma^n[D^m,D_n]\chi_m
-e(1-\bk)\,\bar\sigma^mF_{mn}\chi^n
=
-\,i\,\bk\,\frac{e}{M}D^m\!\left((F_+)_{mn}\bar\lambda^n\right).
\label{eq:beta_left_divergence_sigma_pair_clean}
\end{align}

The commutator acting on a vector--spinor reads
\begin{equation}
[D_m,D_n]\chi_p
=
-\,R_{mnp}{}^{q}\chi_q
+\frac14\,R_{mnab}\,\sigma^{ab}\chi_p
+i e F_{mn}\chi_p.
\label{eq:beta_commutator_chi_clean}
\end{equation}
Its purely gravitational contribution reorganizes into the Einstein-tensor channel,
\begin{equation}
i\,\bar\sigma^n[D^m,D_n]\chi_m\Big|_{\rm grav}
=
\frac{i}{2}\,\bar\sigma^m G_{mn}\chi^n.
\label{eq:beta_left_grav_channel_clean}
\end{equation}
The electromagnetic commutator term combines with the explicit linear Pauli terms already present in
\eqref{eq:beta_left_divergence_sigma_pair_clean}.

We restrict to the constant-field sector,
\begin{equation}
\nabla_m F^{mn}=0,
\qquad
\nabla_{[m}F_{np]}=0,
\label{eq:beta_constantF_background_clean}
\end{equation}
so that \(D^m(F_+)_{mn}=0\). The right-hand side of
\eqref{eq:beta_left_divergence_sigma_pair_clean} reduces to
\begin{equation}
-\,i\,\bk\,\frac{e}{M}(F_+)_{mn}D^m\bar\lambda^n.
\label{eq:beta_left_Pauli_div_target_clean}
\end{equation}

To evaluate this contraction, we rewrite \eqref{eq:beta_family_red_2_clean} as
\begin{equation}
i\,\sigma^p D_p\bar\lambda_n
=
-\,M\,\chi_n
-\,i\,\frac{e}{M}\,\cP_{nq}\chi^q,
\qquad
\cP_{nq}\equiv 2(1-\bk)F_{nq}+\bk(F_-)_{nq}.
\label{eq:beta_right_eq_rewritten_clean}
\end{equation}
Using \(2F=(F_+)+(F_-)\), this becomes
\begin{equation}
\cP_{nq}=(1-\bk)(F_+)_{nq}+(F_-)_{nq}.
\label{eq:beta_P_tensor_split_clean}
\end{equation}

The reduced equations determine only \(\sigma^p D_p\bar\lambda_n\). 
The contraction \((F_+)^{mn}D_m\bar\lambda_n\) is extracted by differentiating the identity
\begin{equation}
\sigma_m(F_+)^{mn}\bar\lambda_n=0,
\label{eq:beta_chiral_identity_clean}
\end{equation}
which follows from the primary constraint. In the constant-field sector this gives
\begin{equation}
\sigma_m(F_+)^{mn}D_p\bar\lambda_n=0.
\label{eq:beta_chiral_identity_diff_clean}
\end{equation}
Contracting with \(\bar\sigma^p\) and using standard Weyl identities yields
\begin{equation}
(F_+)^{pn}D_p\bar\lambda_n
+
2\,\bar\sigma^p{}_m(F_+)^{mn}D_p\bar\lambda_n
=0.
\label{eq:beta_F_D_identity_clean}
\end{equation}
Combining this relation with \eqref{eq:beta_right_eq_rewritten_clean}, one finds
\begin{equation}
i(F_+)^{pn}D_p\bar\lambda_n
=
-\frac{M}{2}\,\bar\sigma_m(F_+)^{mn}\chi_n
-\frac{i e}{2M}\,\bar\sigma_m(F_+)^{mn}\cP_{nq}\chi^q.
\label{eq:beta_projected_pauli_div_clean}
\end{equation}

The first term is linear in \(F\) and combines with existing contributions. 
The genuinely new algebraic term is quadratic in \(F\),
\begin{equation}
\cQ^{(L)}
=
i\bk\,\frac{e^2}{2M^2}\,
(F_+)^{mn}\bar\sigma_m\,\cP_{nq}\chi^q.
\label{eq:beta_left_Q_start_clean}
\end{equation}
Using \eqref{eq:beta_P_tensor_split_clean}, the chiral-pure term proportional to
\((F_+)^{mn}(F_+)_{nq}\) reduces to the primary constraint and does not contribute independently.
The mixed product yields
\begin{equation}
(F_+)^{mn}(F_-)_{nq}=2\,T^{(F)m}{}_{q},
\label{eq:beta_mixed_identity_clean}
\end{equation}
and therefore
\begin{equation}
\cQ^{(L)}
=
i\bk\,\frac{e^2}{M^2}\,
\bar\sigma^mT^{(F)}_{mn}\chi^n
+\text{(constraint- or eom-controlled terms)}.
\label{eq:beta_left_Q_stress_clean}
\end{equation}

The divergence generates an \emph{algebraic} rank-two structure, which is the crucial 
difference from the flat-space analysis. While the linear terms in \(F\) combine with contributions already
present in \eqref{eq:beta_left_divergence_sigma_pair_clean}, the quadratic term
\eqref{eq:beta_left_Q_stress_clean} produces a genuinely new channel proportional to the
Maxwell stress tensor \(T^{(F)}_{mn}\).

This is the crucial difference with the flat-space analysis: the obstruction is no longer
purely differential, but contains an algebraic tensor contribution. As we will see below,
this allows a nontrivial matching with the gravitational sector.

Substituting into \eqref{eq:beta_left_divergence_sigma_pair_clean}, we obtain
\begin{equation}
i\,\bar\sigma^n D_n \fS^{(L)}
+M\,\fS^{(R)}
+\frac{i}{2}\,\bar\sigma^m G_{mn}\chi^n
-\,i\bk\,\frac{e^2}{M^2}\,\bar\sigma^mT^{(F)}_{mn}\chi^n
=0,
\label{eq:beta_left_closure_clean}
\end{equation}
up to terms proportional to the primary constraints or to the equations of motion.

The right-handed sector follows by conjugation,
\begin{equation}
i\,\sigma^n D_n \fS^{(R)}
+M\,\fS^{(L)}
+\frac{i}{2}\,\sigma^m G_{mn}\bar\lambda^n
-\,i\bk\,\frac{e^2}{M^2}\,\sigma^mT^{(F)}_{mn}\bar\lambda^n
=0.
\label{eq:beta_right_closure_clean}
\end{equation}

These relations show that no new independent differential constraint appears at this
stage. Instead, the next step of the chain reorganizes the secondary constraints through
an Einstein--Maxwell dressing. The obstruction is therefore not eliminated, but reshaped
into a combination of geometric and matter tensor structures.

This motivates the definitions
\begin{align}
\fZ^{(L)}_{\alpha}
&=
\fS^{(L)}_{\alpha}
+\frac{i}{2M}\,\sigma^m_{\alpha\dot\alpha}\,G_{mn}\,\bar\lambda^{n\dot\alpha}
-\frac{i\bk e^2}{M^3}\,
\sigma^m_{\alpha\dot\alpha}\,T^{(F)}_{mn}\,\bar\lambda^{n\dot\alpha},
\\
\fZ^{(R)}_{\dot\alpha}
&=
\fS^{(R)}_{\dot\alpha}
+\frac{i}{2M}\,\bar\sigma^m_{\dot\alpha\alpha}\,G_{mn}\,\chi^{n\alpha}
-\frac{i\bk e^2}{M^3}\,
\bar\sigma^m_{\dot\alpha\alpha}\,T^{(F)}_{mn}\,\chi^{n\alpha},
\end{align}
for which the closure takes the compact form
\begin{equation}
i\,\bar\sigma^n D_n \fS^{(L)} + M\,\fZ^{(R)} = 0,
\qquad
i\,\sigma^n D_n \fS^{(R)} + M\,\fZ^{(L)} = 0,
\end{equation}
up to terms already controlled by lower levels of the chain.

The reduced system therefore propagates consistently with the constraint chain
\begin{equation}
\bar\sigma^m\chi_m=0,
\quad
\sigma^m\bar\lambda_m=0,
\quad
\fS^{(L)}=0,
\quad
\fS^{(R)}=0,
\quad
\fZ^{(L)}=0,
\quad
\fZ^{(R)}=0.
\end{equation}

\subsection{Einstein--Maxwell reduction and tuned locus}

The structure obtained above shows that the obstruction generated by the divergence
naturally decomposes into two algebraic rank-two channels: a gravitational contribution
proportional to the Einstein tensor and an electromagnetic contribution proportional to
the Maxwell stress tensor. Closure is therefore possible only if these two contributions
can be matched.

On Einstein--Maxwell backgrounds,
\begin{equation}
G_{mn}=\kappa^2T^{(F)}_{mn}-\Lambda g_{mn},
\end{equation}
the dressed constraints reduce to
\begin{align}
\fZ^{(L)}_{\alpha}
&=
\fS^{(L)}_{\alpha}
-\frac{i\Lambda}{2M}\,\sigma^m_{\alpha\dot\alpha}\bar\lambda_m{}^{\dot\alpha}
+i\left(
\frac{\kappa^2}{2M}
-\frac{\bk e^2}{M^3}
\right)\sigma^m_{\alpha\dot\alpha}T^{(F)}_{mn}\bar\lambda^{n\dot\alpha},
\\
\fZ^{(R)}_{\dot\alpha}
&=
\fS^{(R)}_{\dot\alpha}
-\frac{i\Lambda}{2M}\,\bar\sigma^m_{\dot\alpha\alpha}\chi_m{}^{\alpha}
+i\left(
\frac{\kappa^2}{2M}
-\frac{\bk e^2}{M^3}
\right)\bar\sigma^m_{\dot\alpha\alpha}T^{(F)}_{mn}\chi^{n\alpha}.
\end{align}
The cosmological term vanishes on the primary constraint surface, so the nontrivial obstruction is controlled entirely by the Maxwell stress tensor.

For \(\bk\neq0\), closure of the constraint chain requires cancellation of the
Maxwell stress-tensor channel. This fixes the relation
\begin{equation}
\frac{\kappa^2}{2M}=\frac{\bk e^2}{M^3}
\qquad\Longleftrightarrow\qquad
M^2=\frac{2\bk e^2}{\kappa^2}.
\end{equation}

At \(\bk=1\), this reproduces the standard \(\cN=2\) supergravity locus. 
At \(\bk=0\), the quadratic Maxwell contribution is absent and no matching condition
arises, corresponding to the decoupled regime where gravity does not participate in the
constraint algebra.

The tuned locus therefore emerges as the unique condition under which the
Einstein-tensor and Maxwell stress-tensor channels match in the constraint chain.

\subsection{Neutral limit and relation to Buchdahl's condition}
\label{sec:neutral_limit_buchdahl}

We now switch off the electromagnetic background and isolate the purely
gravitational content of the spin-\(\tfrac32\) Fierz--Pauli chain. 
This limit serves two purposes. First, it shows that the Einstein--Maxwell
analysis reduces to the standard Einstein-space condition in the absence of
electromagnetic fields. Second, it clarifies the relation with the classical
result of Buchdahl.

Setting
\begin{equation}
F_{\mu\nu}=0,
\end{equation}
all Pauli terms disappear and the reduced system collapses to the neutral
covariant Fierz--Pauli equations
\begin{equation}
\mathcal R^\mu(\Psi)
\equiv
\gamma^{\mu\nu\rho}\nabla_\nu\Psi_\rho
+M\,\gamma^{\mu\nu}\Psi_\nu
=0,
\label{eq:neutral_RS_system}
\end{equation}
together with the primary constraint
\begin{equation}
\gamma^\mu\Psi_\mu=0.
\label{eq:neutral_primary_constraint}
\end{equation}

Taking the covariant divergence of \eqref{eq:neutral_RS_system} and using the
primary constraint together with the commutator of covariant derivatives on a
vector--spinor, one finds, up to terms already controlled by lower-spin
constraints,
\begin{equation}
\mathcal C(\Psi)\big|_{F=0}
\;\sim\;
G_{\mu\nu}\,\gamma^\mu\Psi^\nu.
\label{eq:neutral_Einstein_channel}
\end{equation}

In the neutral case there is no additional matter contribution that could cancel
this term. Closure of the Fierz--Pauli chain therefore requires the background
to satisfy
\begin{equation}
G_{\mu\nu}+\Lambda g_{\mu\nu}=0,
\label{eq:neutral_Einstein_condition}
\end{equation}
or equivalently
\begin{equation}
R_{\mu\nu}=\Lambda g_{\mu\nu}.
\label{eq:neutral_Einstein_space}
\end{equation}
Thus, consistency of the covariant spin-\(\tfrac32\) system selects Einstein
backgrounds.

The above result is the neutral limit of the mechanism identified in the
Einstein--Maxwell case. There, the divergence algebra produces an
Einstein-tensor channel which, upon using Einstein's equations, becomes a
Maxwell stress-tensor contribution. Closure is then possible because the Pauli
sector reproduces the same rank-two structure. When \(F_{\mu\nu}=0\), this
matching mechanism is absent, and one is left with the pure Einstein condition
\eqref{eq:neutral_Einstein_space}.

It is instructive to compare this conclusion with the classical analysis of
Buchdahl~\cite{Buchdahl:1958xv}. In modern notation, Buchdahl showed that the naive
covariantization of the massive spin-\(\tfrac32\) equations leads, upon taking
the divergence and using \(\gamma\!\cdot\!\Psi=0\), to the condition
\begin{equation}
R_{\mu\nu}\,\gamma^\mu\Psi^\nu = 0,
\label{eq:buchdahl_condition}
\end{equation}
which must hold for arbitrary \(\Psi_\mu\). This implies that the background
satisfies
\begin{equation}
R_{\mu\nu} \propto g_{\mu\nu},
\end{equation}
i.e. that it is an Einstein space.

The physical content of \eqref{eq:neutral_Einstein_space} is therefore identical
to Buchdahl's result. The difference lies in the formulation. Buchdahl derives
\eqref{eq:buchdahl_condition} as a compatibility condition of the covariantized
equations. Here the same requirement appears as closure of the Fierz--Pauli
constraint chain: starting from the primary constraint, one derives the
secondary divergence and finds that the next obstruction is precisely the
Einstein-tensor term \eqref{eq:neutral_Einstein_channel}. The present approach
thus reorganizes the same condition within the constraint-algebra framework.

The conclusion can be summarized as follows: In the neutral limit \(F_{\mu\nu}=0\), closure of the covariant
spin-\(\tfrac32\) Fierz--Pauli chain reproduces the Buchdahl
consistency condition that the background be Einstein. The result is the same
in substance, while the derivation is organized here in terms of the constraint
chain.

%
%
%
%
%
%
%

%
%
%
%
%
%
%

%
%
%
%
%
%
%

\section{Exact constant-field second-order equations on Einstein--Maxwell backgrounds}
\label{sec:constant_field_second_order}

In this section we derive the exact second-order equations obeyed by the reduced
\(\beta_\kappa\)-family on Einstein--Maxwell backgrounds with covariantly constant
field strength. The purpose is twofold. First, the second-order form makes the linear Zeeman
couplings manifest and therefore allows an unambiguous identification of the
gyromagnetic ratio. Second, it separates the quadratic \(F^2\) structures into
two distinct contributions: a tensorial channel proportional to the Maxwell
stress tensor and a scalar chiral invariant. As we will see, the former is
precisely the channel responsible for the closure condition derived in
section~\ref{sec:beta_kappa_closure_core}, while the latter remains as a genuine
residual deformation of the wave operator.

Throughout this section the background satisfies
\begin{equation}
G_{mn}+\Lambda g_{mn}=\kappa^2 T^{(F)}_{mn},
\qquad
T^{(F)}_{mn}=-F_{mp}F_n{}^{p}+\frac14 g_{mn}F^2,
\qquad
\nabla_\ell F_{mn}=0.
\label{eq:g_curved_EM_background}
\end{equation}
In four dimensions the Maxwell stress tensor is traceless, so
\begin{equation}
R=4\Lambda,
\qquad
R_{mn}=\Lambda g_{mn}+\kappa^2 T^{(F)}_{mn}.
\label{eq:g_curved_Ricci_EM}
\end{equation}

We start from the reduced first-order system
\begin{align}
i\,\bar\sigma^n D_n \chi_m + M\,\bar\lambda_m
&=
-\,i\,\beta_\kappa\,\frac{e}{M}(F_+)_{mn}\bar\lambda^n,
\label{eq:g_curved_red1}
\\[1mm]
i\,\sigma^n D_n \bar\lambda_m + M\,\chi_m
&=
-\,i\,\frac{e}{M}\Bigl[2(1-\beta_\kappa)F_{mn}+\beta_\kappa(F_-)_{mn}\Bigr]\chi^n,
\label{eq:g_curved_red2}
\end{align}
together with the primary constraints
\begin{equation}
\bar\sigma^m\chi_m=0,
\qquad
\sigma^m\bar\lambda_m=0.
\label{eq:g_curved_primary}
\end{equation}

For notational convenience we introduce
\begin{equation}
A_m{}^{n}\equiv \beta_\kappa\,\frac{e}{M}(F_+)_m{}^{n},
\qquad
B_m{}^{n}\equiv \frac{e}{M}\Bigl[2(1-\beta_\kappa)F_m{}^{n}+\beta_\kappa(F_-)_m{}^{n}\Bigr].
\label{eq:g_curved_A_B}
\end{equation}
The system reads
\begin{equation}
i\,\bar\sigma\!\cdot\!D\,\chi_m+(M\delta_m{}^n+iA_m{}^{n})\bar\lambda_n=0,
\qquad
i\,\sigma\!\cdot\!D\,\bar\lambda_m+(M\delta_m{}^n+iB_m{}^{n})\chi_n=0.
\label{eq:g_curved_matrix_system}
\end{equation}
A useful identity is
\begin{equation}
A+B=\frac{2e}{M}F.
\label{eq:g_curved_AplusB}
\end{equation}

\subsection{Curved-space constant-field second-order equations}

The curved-space squaring identities for Weyl vector--spinors are
\begin{equation}
(\sigma\!\cdot\!D)(\bar\sigma\!\cdot\!D)\chi_m
=
D^2\chi_m
+ie\,\sigma^{ab}F_{ab}\chi_m
+\Bigl(R_m{}^{n}-\frac14 R\,\delta_m{}^{n}\Bigr)\chi_n,
\label{eq:g_curved_weyl_square_left}
\end{equation}
and
\begin{equation}
(\bar\sigma\!\cdot\!D)(\sigma\!\cdot\!D)\bar\lambda_m
=
D^2\bar\lambda_m
+ie\,\bar\sigma^{ab}F_{ab}\bar\lambda_m
+\Bigl(R_m{}^{n}-\frac14 R\,\delta_m{}^{n}\Bigr)\bar\lambda_n.
\label{eq:g_curved_weyl_square_right}
\end{equation}
Using \eqref{eq:g_curved_Ricci_EM}, this becomes
\begin{equation}
R_m{}^{n}-\frac14R\,\delta_m{}^{n}
=
\kappa^2 T^{(F)}_m{}^{n}.
\label{eq:g_curved_Ricci_traceless}
\end{equation}

Thus, on Einstein--Maxwell backgrounds with covariantly constant field strength,
all curvature corrections entering the second-order operator are encoded in the
Maxwell stress tensor. No independent Weyl or Ricci-trace structures appear in
this sector.

Acting with \(i\,\sigma\!\cdot\!D\) on \eqref{eq:g_curved_matrix_system}, and using
\(\nabla F=0\), we obtain
\begin{equation}
(\sigma\!\cdot\!D)(\bar\sigma\!\cdot\!D)\chi_m
+\bigl[(M+iA)(M+iB)\bigr]_m{}^{r}\chi_r=0.
\label{eq:g_curved_chi_second_matrix}
\end{equation}
Expanding,
\begin{equation}
(M+iA)(M+iB)=M^2+iM(A+B)-AB.
\end{equation}
The linear term gives
\begin{equation}
iM(A+B)=2ie\,F,
\end{equation}
while
\begin{equation}
AB
=
\beta_\kappa\frac{e^2}{M^2}\Bigl[(1-\beta_\kappa)F_+^2+F_+F_-\Bigr].
\end{equation}
Using
\begin{equation}
(F_+F_-)_m{}^{n}=2T^{(F)}_m{}^{n},
\qquad
(F_+^2)_m{}^{n}
=
-\frac14(F_+)_{ab}(F_+)^{ab}\,\delta_m{}^{n},
\end{equation}
we obtain
\begin{equation}
AB_m{}^{n}
=
\frac{2\beta_\kappa e^2}{M^2}T^{(F)}_m{}^{n}
-\frac{\beta_\kappa(1-\beta_\kappa)e^2}{4M^2}(F_+)^2\delta_m{}^{n}.
\end{equation}

Substituting, we find
\begin{equation}
\begin{aligned}
0={}&(D^2+M^2)\chi_m
+ie\,\sigma^{ab}F_{ab}\chi_m
+2ie\,F_{mn}\chi^n \\
&+\left(\kappa^2-\frac{2\beta_\kappa e^2}{M^2}\right)T^{(F)}_{mn}\chi^n \\
&+\frac{\beta_\kappa(1-\beta_\kappa)e^2}{4M^2}(F_+)^2\chi_m .
\end{aligned}
\label{eq:g_curved_chi_second}
\end{equation}

The same computation gives
\begin{equation}
\begin{aligned}
0={}&(D^2+M^2)\bar\lambda_m
+ie\,\bar\sigma^{ab}F_{ab}\bar\lambda_m
+2ie\,F_{mn}\bar\lambda^n \\
&+\left(\kappa^2-\frac{2\beta_\kappa e^2}{M^2}\right)T^{(F)}_{mn}\bar\lambda^n \\
&+\frac{\beta_\kappa(1-\beta_\kappa)e^2}{4M^2}(F_-)^2\bar\lambda_m .
\end{aligned}
\label{eq:g_curved_lbar_second}
\end{equation}

The structure makes the mechanism of section~\ref{sec:beta_kappa_closure_core}
explicit: the quadratic contribution splits into a tensorial stress-tensor channel
and a scalar chiral invariant.

The structure explicitly demonstrates the mechanism of Section~\ref{sec:beta_kappa_closure_core}: 
the quadratic sector separates into a tensorial channel 
(proportional to the Maxwell stress tensor) and a scalar chiral invariant.

On the tuned locus
\begin{equation}
M^2=\frac{2\beta_\kappa e^2}{\kappa^2},
\end{equation}
the stress-tensor channel cancels identically. This is the second-order
realization of the closure condition: the same rank-two tensor appears from the
commutator algebra and from the Pauli sector.

\subsection{Gyromagnetic ratio and chiral scalar term}

The exact tuned equations make the linear electromagnetic couplings manifest. In
the \(\chi_m\) equation they are
\begin{equation}
ie\,\sigma^{ab}F_{ab}\chi_m,
\qquad
2ie\,F_{mn}\chi^n,
\end{equation}
and similarly, in the \(\bar\lambda_m\) equation,
\begin{equation}
ie\,\bar\sigma^{ab}F_{ab}\bar\lambda_m,
\qquad
2ie\,F_{mn}\bar\lambda^n.
\end{equation}
These terms act on distinct indices:  the first is a 
spin-\(\tfrac12\) Zeeman operator acting on the Weyl-spinor index,
while the second is a spin-1 Zeeman operator on the vector index.~\cite{Benakli:2025xao}.
Comparing with the universal second-order form, one reads off
\begin{equation}
g=2.
\end{equation}

This result is independent of \(\beta_\kappa\). The interpolation parameter
affects only the quadratic \(F^2\) sector and plays no role in the linear Zeeman
couplings.

The residual quadratic terms are proportional to the chiral invariants
\begin{equation}
(F_\pm)^2 \equiv (F_\pm)_{ab}(F_\pm)^{ab}
=
2\bigl(F_{ab}F^{ab}\pm i\,F_{ab}\widetilde F^{ab}\bigr).
\end{equation}
Thus, for all real \(\beta_\kappa\) except the two endpoints \(\beta_\kappa=0\)
and \(\beta_\kappa=1\), the tuned second-order equations contain a chiral scalar
\(F^2\) deformation proportional to \(\beta_\kappa(1-\beta_\kappa)\). Whenever
\begin{equation}
F\widetilde F\neq0,
\end{equation}
these chiral invariants are complex. In that case the exact constant-field wave
operator is no longer manifestly real or Hermitian.

This observation does not by itself establish the presence of ghosts. A ghost
diagnosis requires a direct analysis of the kinetic form and of the norm of the
propagating modes. What can be stated rigorously is more limited but still
significant: for all real \(\beta_\kappa\neq0,1\), backgrounds with
\(F\widetilde F\neq0\) induce a complex scalar deformation of the exact
constant-field wave operator. One should then generically expect complex
frequencies or non-oscillatory growth/decay for some modes.

The key conclusion is therefore the following: The closure and universal value \(g=2\) does 
not guarantee consistency of the Fierz--Pauli system. The distinction
between the interpolating theories appears at quadratic order.

\subsection{Rigidity within simple first-order counterterms}

We now ask whether the chiral scalar obstruction can be removed by deforming the
reduced first-order system. We restrict attention to local two-derivative
deformations that preserve the structure of the reduced equations and leave the
linear \(g=2\) Zeeman sector unchanged.

A scalar \(\mathcal O(F^2)\) deformation shifts both chiral sectors equally.
It therefore cannot cancel the chiral terms proportional to \((F_+)^2\) and
\((F_-)^2\), which are generically different whenever \(F\widetilde F\neq0\).

A tensor \(\mathcal O(F^2)\) deformation does not improve the situation.
In the relevant vector-index channel, only its trace part can contribute
directly to the scalar obstruction, and this reduces the problem to the same
failed scalar case.

Finally, linear \(\mathcal O(F)\) deformations preserving the \(g=2\) Zeeman
sector must satisfy
\begin{equation}
\delta A+\delta B=0.
\end{equation}
But the resulting corrections to the quadratic sector necessarily generate
incompatible conditions: canceling the unwanted chiral \((F_+)^2\) term
introduces either a new \((F_-)^2\) contribution or a shift in the
mixed \(F_+F_-\) stress-tensor channel.

We conclude: Within local two-derivative first-order deformations preserving the structure of
the reduced system and the \(g=2\) sector, the chiral scalar obstruction is
rigid. Only the endpoints \(\beta_\kappa=0\) and \(\beta_\kappa=1\) avoid it.

It is important to stress the status of this result. The chiral scalar term is an
exact feature of the constant-field local two-derivative system studied in this
section. It is not a higher-derivative remainder in a derivative expansion,
since the background satisfies \(\nabla_\ell F_{mn}=0\) throughout. In
particular, the term is part of the exact second-order operator obtained by
squaring the first-order Fierz--Pauli system in the constant-field regime.

One may then ask whether this scalar term could be removed by a local field
redefinition or by simple local counterterms. Within the restricted class of
local first-order deformations analyzed below, the answer is negative: the term
is rigid. The point is not that every conceivable completion is excluded, but
that this chiral scalar structure is not a derivative-expansion artifact of the
constant-field computation, and it cannot be removed inside the same minimal
local two-derivative Fierz--Pauli framework without changing other essential
parts of the operator.

%
%
%
%
%
%
%

%
%
%
%
%
%
%

%
%
%
%
%
%
%

%
%
%
%
%
%
%

\section{Non-constant electromagnetic backgrounds and the \texorpdfstring{$\partial F$}{dF} obstruction}
\label{sec:nonconstant_F}

We now relax the assumption of covariantly constant field strength and allow
\(F_{mn}\) to vary in spacetime. This provides a stronger test of the reduced
\(\beta_\kappa\)-family, since the constant-field analysis does not probe the terms generated when
derivatives act on the Pauli couplings themselves.

The outcome is straightforward: the secondary divergences remain unchanged, but the 
next step in the chain acquires new \(\nabla F\) terms. For generic \(\beta_\kappa\), these terms are
not absorbed by the dressed Einstein--Maxwell obstruction identified in the constant-field analysis. At
the symmetric endpoint \(\beta_\kappa=1\), however, the naked \(\partial F\) term cancels exactly,
and only the physical Maxwell current remains.

We keep the reduced equations
\begin{align}
i\,\bar\sigma^n D_n \chi_m + M\,\bar\lambda_m
&=
-\,i\,\beta_\kappa\,\frac{e}{M}\,(F_+)_{mn}\,\bar\lambda^n,
\label{eq:nonconstF_red_eq1}
\\[1mm]
i\,\sigma^n D_n \bar\lambda_m + M\,\chi_m
&=
-\,i\,\frac{e}{M}
\Bigl[
2(1-\beta_\kappa)\,F_{mn}
+\beta_\kappa\,(F_-)_{mn}
\Bigr]\chi^n,
\label{eq:nonconstF_red_eq2}
\end{align}
supplemented by the primary constraints
\begin{equation}
\bar\sigma^m\chi_m=0,
\qquad
\sigma^m\bar\lambda_m=0.
\label{eq:nonconstF_primary}
\end{equation}
We also define the Maxwell current
\begin{equation}
j_n\equiv \nabla^mF_{mn}.
\label{eq:maxwell_current_def}
\end{equation}
Since the dual Bianchi identity implies \(\nabla^m\widetilde F_{mn}=0\), one has
\begin{equation}
\nabla^m(F_\pm)_{mn}=j_n.
\label{eq:Fpm_divergence_equals_j}
\end{equation}

The derivation of the secondary divergences uses only the \(\sigma\)-traces of
\eqref{eq:nonconstF_red_eq1}--\eqref{eq:nonconstF_red_eq2}. No derivative acts on \(F_{mn}\) at
that stage, so the secondary pair is unchanged:
\begin{equation}
\Sigma^{(L)}_\alpha \equiv D^m\chi_{m\alpha}=0,
\qquad
\Sigma^{(R)}_{\dot\alpha}
\equiv
D^m\bar\lambda_{m\dot\alpha}
+\frac{e}{M}(1-\beta_\kappa)\,\bar\sigma^m_{\dot\alpha\alpha}F_{mn}\chi^{n\alpha}=0.
\label{eq:nonconstF_sigma_pair}
\end{equation}
Equivalently,
\begin{equation}
D\!\cdot\!\chi=\Sigma^{(L)},
\qquad
D\!\cdot\!\bar\lambda
=
\Sigma^{(R)}
-\frac{e}{M}(1-\beta_\kappa)\,\bar\sigma^mF_{mq}\chi^q.
\label{eq:nonconstF_Ddot_substitution_pair}
\end{equation}

We begin with the divergence of the left reduced equation. Using \(D_m\bar\sigma^n=0\), one finds
\begin{equation}
i\,\bar\sigma^n D_n\Sigma^{(L)}
+i\,\bar\sigma^n[D^m,D_n]\chi_m
+M\,D\!\cdot\!\bar\lambda
=
-\,i\,\beta_\kappa\,\frac{e}{M}\,
D^m\!\left((F_+)_{mn}\bar\lambda^n\right).
\label{eq:nonconstF_div_left_start}
\end{equation}
Expanding the Pauli divergence on the right-hand side gives
\begin{equation}
D^m\!\left((F_+)_{mn}\bar\lambda^n\right)
=
(\nabla^m(F_+)_{mn})\bar\lambda^n
+
(F_+)_{mn}D^m\bar\lambda^n
=
j_n\bar\lambda^n
+
(F_+)_{mn}D^m\bar\lambda^n.
\label{eq:nonconstF_left_pauli_div}
\end{equation}
Using also \eqref{eq:nonconstF_Ddot_substitution_pair}, we obtain
\begin{equation}
\begin{aligned}
i\,\bar\sigma^n D_n\Sigma^{(L)}
+M\,\Sigma^{(R)}
+i\,\bar\sigma^n[D^m,D_n]\chi_m
-e(1-\beta_\kappa)\,\bar\sigma^mF_{mn}\chi^n
={}&-\,i\,\beta_\kappa\,\frac{e}{M}\,j_n\bar\lambda^n \\
&-\,i\,\beta_\kappa\,\frac{e}{M}(F_+)_{mn}D^m\bar\lambda^n .
\end{aligned}
\label{eq:nonconstF_left_before_reduction}
\end{equation}

The second term on the right-hand side is the same projected Pauli divergence that appeared in the
constant-field analysis. Its reduction is unchanged and therefore contributes to the same dressed
object as before. The only genuinely new term in the left sector is the current source
\(-i\beta_\kappa(e/M)\,j_n\bar\lambda^n\). Thus, in the same notation as in the
constant-field section,
\begin{equation}
i\,\bar\sigma^n D_n \Sigma^{(L)} + M\,\mathfrak Z^{(R)}
=
-\,i\,\beta_\kappa\,\frac{e}{M}\,j_n\bar\lambda^n,
\label{eq:nonconstF_compact_pair_L}
\end{equation}
where \(\mathfrak Z^{(L,R)}\) are the same dressed Einstein--Maxwell combinations as in the
constant-field case.

The decisive effect appears in the divergence of the right reduced equation. Starting from
\eqref{eq:nonconstF_red_eq2}, define
\begin{equation}
\mathcal P_{pq}\equiv 2(1-\beta_\kappa)F_{pq}+\beta_\kappa(F_-)_{pq}.
\label{eq:nonconstF_P_def}
\end{equation}
Then
\begin{equation}
i\,D^p\!\left(\sigma^n D_n\bar\lambda_p\right)
+M\,D^p\chi_p
=
-\,i\,\frac{e}{M}\,
D^p\!\left(\mathcal P_{pq}\chi^q\right).
\label{eq:nonconstF_div_right_start}
\end{equation}
Using \(D\!\cdot\!\chi=\Sigma^{(L)}\) and splitting the derivative on the left gives
\begin{equation}
i\,\sigma^n D_n(D\!\cdot\!\bar\lambda)
+i\,\sigma^n[D^p,D_n]\bar\lambda_p
+M\,\Sigma^{(L)}
=
-\,i\,\frac{e}{M}\,
D^p\!\left(\mathcal P_{pq}\chi^q\right).
\label{eq:nonconstF_right_master_before_sub}
\end{equation}
Now substitute
\begin{equation}
D\!\cdot\!\bar\lambda
=
\Sigma^{(R)}
-\frac{e}{M}(1-\beta_\kappa)\,\bar\sigma^mF_{mq}\chi^q.
\label{eq:nonconstF_Ddotlambda_sub}
\end{equation}
This is precisely where the non-constant-field effect enters. Applying \(i\sigma^nD_n\) gives
\begin{align}
i\,\sigma^n D_n(D\!\cdot\!\bar\lambda)
&=
i\,\sigma^n D_n \Sigma^{(R)}
-\,i\,\frac{e}{M}(1-\beta_\kappa)\,
\sigma^n\bar\sigma^m(\nabla_n F_{mq})\chi^q
\nonumber\\
&\qquad
-\,i\,\frac{e}{M}(1-\beta_\kappa)\,
\sigma^n\bar\sigma^mF_{mq}D_n\chi^q .
\label{eq:nonconstF_DnDdotlambda}
\end{align}
Using
\begin{equation}
\sigma^n\bar\sigma^m = g^{nm}+2\sigma^{nm},
\label{eq:clifford_sigma_barsigma_nonconstF}
\end{equation}
the derivative-of-\(F\) term becomes
\begin{equation}
\sigma^n\bar\sigma^m(\nabla_n F_{mq})
=
\nabla^m F_{mq}
+2\sigma^{nm}(\nabla_n F_{mq})
=
j_q+2\sigma^{nm}(\nabla_n F_{mq}).
\label{eq:nablaF_decomposition}
\end{equation}
Hence
\begin{align}
i\,\sigma^n D_n(D\!\cdot\!\bar\lambda)
&=
i\,\sigma^n D_n \Sigma^{(R)}
-\,i\,\frac{e}{M}(1-\beta_\kappa)\,j_q\chi^q
\nonumber\\
&\qquad
-\,2i\,\frac{e}{M}(1-\beta_\kappa)\sigma^{nm}(\nabla_n F_{mq})\chi^q
-\,i\,\frac{e}{M}(1-\beta_\kappa)\sigma^n\bar\sigma^mF_{mq}D_n\chi^q.
\label{eq:nonconstF_DnDdotlambda_final}
\end{align}

We next evaluate the divergence of the Pauli term on the right-hand side of
\eqref{eq:nonconstF_right_master_before_sub}:
\begin{equation}
D^p\!\left(\mathcal P_{pq}\chi^q\right)
=
(\nabla^p\mathcal P_{pq})\chi^q
+
\mathcal P_{pq}D^p\chi^q.
\label{eq:nonconstF_right_pauli_div}
\end{equation}
Using \(\nabla^p(F_-)_{pq}=j_q\), one finds
\begin{equation}
\nabla^p\mathcal P_{pq}
=
2(1-\beta_\kappa)\,j_q+\beta_\kappa\,j_q
=
(2-\beta_\kappa)\,j_q.
\label{eq:nonconstF_P_divergence}
\end{equation}
Therefore the right-hand side contributes
\begin{equation}
-\,i\,\frac{e}{M}(2-\beta_\kappa)\,j_q\chi^q
-\,i\,\frac{e}{M}\mathcal P_{pq}D^p\chi^q.
\label{eq:nonconstF_right_rhs_expanded}
\end{equation}

Substituting \eqref{eq:nonconstF_DnDdotlambda_final} and
\eqref{eq:nonconstF_right_rhs_expanded} into
\eqref{eq:nonconstF_right_master_before_sub}, the terms involving \(F_{mq}D_n\chi^q\), together
with the commutator term \(i\,\sigma^n[D^p,D_n]\bar\lambda_p\), reduce exactly as in the
constant-field analysis to the same dressed object \(\mathfrak Z^{(L)}\). The genuinely new
terms generated by non-constant fields are therefore the current and gradient terms written
explicitly above.

The coefficient of the current term is universal. Indeed, the contribution coming from
\eqref{eq:nonconstF_DnDdotlambda_final} is proportional to \(1-\beta_\kappa\), while the one from
\eqref{eq:nonconstF_right_rhs_expanded} is proportional to \(2-\beta_\kappa\); their difference is
independent of \(\beta_\kappa\). One thus obtains
\begin{equation}
i\,\sigma^n D_n \Sigma^{(R)} + M\,\mathfrak Z^{(L)}
=
-\,i\,\frac{e}{M}\,j_n\chi^n
+\,2i\,\frac{e}{M}(1-\beta_\kappa)\sigma^{nm}(\nabla_n F_{mq})\chi^q.
\label{eq:nonconstF_compact_pair_R}
\end{equation}

The dressed Einstein--Maxwell combinations themselves are unchanged from the constant-field
analysis:
\begin{equation}
\mathfrak Z^{(L)}_{\alpha}
=
\Sigma^{(L)}_{\alpha}
+\frac{i}{2M}\,\sigma^m_{\alpha\dot\alpha}G_{mn}\bar\lambda^{n\dot\alpha}
-\frac{i\beta_\kappa e^2}{M^3}\,
\sigma^m_{\alpha\dot\alpha}T^{(F)}_{mn}\bar\lambda^{n\dot\alpha},
\label{eq:nonconstF_ZL}
\end{equation}
\begin{equation}
\mathfrak Z^{(R)}_{\dot\alpha}
=
\Sigma^{(R)}_{\dot\alpha}
+\frac{i}{2M}\,\bar\sigma^m_{\dot\alpha\alpha}G_{mn}\chi^{n\alpha}
-\frac{i\beta_\kappa e^2}{M^3}\,
\bar\sigma^m_{\dot\alpha\alpha}T^{(F)}_{mn}\chi^{n\alpha}.
\label{eq:nonconstF_ZR}
\end{equation}
What changes is the propagation system for \(\Sigma^{(L)}\) and \(\Sigma^{(R)}\):
\begin{align}
i\,\bar\sigma^n D_n \Sigma^{(L)} + M\,\mathfrak Z^{(R)}
&=
-\,i\,\beta_\kappa\,\frac{e}{M}\,j_n\bar\lambda^n,
\label{eq:nonconstF_compact_pair_L_repeat}
\\[1mm]
i\,\sigma^n D_n \Sigma^{(R)} + M\,\mathfrak Z^{(L)}
&=
-\,i\,\frac{e}{M}\,j_n\chi^n
+\,2i\,\frac{e}{M}(1-\beta_\kappa)\sigma^{nm}(\nabla_n F_{mq})\chi^q.
\label{eq:nonconstF_compact_pair_R_repeat}
\end{align}

These two equations provide the decisive test. At the symmetric endpoint \(\beta_\kappa=1\), the naked
gradient term vanishes identically, and the only surviving non-constant-\(F\) source is the
physical Maxwell current:
\begin{equation}
i\,\bar\sigma^n D_n \Sigma^{(L)} + M\,\mathfrak Z^{(R)}
=
-\,i\,\frac{e}{M}\,j_n\bar\lambda^n,
\qquad
i\,\sigma^n D_n \Sigma^{(R)} + M\,\mathfrak Z^{(L)}
=
-\,i\,\frac{e}{M}\,j_n\chi^n.
\label{eq:nonconstF_beta1_pair}
\end{equation}
Thus, for a source-free electromagnetic background,
\begin{equation}
j_n=\nabla^mF_{mn}=0,
\label{eq:nonconstF_sourcefree}
\end{equation}
the reduced \(\beta_\kappa=1\) chain remains formally consistent even in the presence of non-constant fields.

For generic \(\beta_\kappa\neq1\), however, the term
\begin{equation}
2i\,\frac{e}{M}(1-\beta_\kappa)\sigma^{nm}(\nabla_n F_{mq})\chi^q
\label{eq:nonconstF_gradient_obstruction}
\end{equation}
survives. This term is not proportional to the Maxwell current and
cannot be absorbed by the dressed constraints 
 \(\Sigma^{(L,R)}\) or \(\mathfrak Z^{(L,R)}\). It therefore represents a genuinely new
lower-spin obstruction. In particular, even if \(j_n=0\), the reduced chain fails unless
\(\beta_\kappa=1\).

We therefore reach a stronger conclusion than in the constant-field analysis. The latter
selected the tuned locus inside the \(\beta_\kappa\)-family. The present non-constant-field
test further selects the symmetric endpoint itself: for \(\beta_\kappa\neq1\), a naked
\(\partial F\) obstruction appears, whereas for \(\beta_\kappa=1\) this term cancels exactly,
leaving only the physical current \(j_\mu\).

It is also useful to state explicitly in what sense this term is ``naked''.
The point is not merely that it contains a derivative of the field strength.
Rather, within the local two-derivative lower-spin chain defined here, it is not
proportional to the Maxwell current \(j_\mu=\nabla^\nu F_{\nu\mu}\), nor is it
absorbed by the previously defined secondary quantities
\(\Sigma^{(L,R)}\) or by their dressed Einstein--Maxwell combinations.
Accordingly, it represents a genuinely new structure relative to the minimal
Fierz--Pauli chain.

Of course, one may enlarge the framework by allowing additional
higher-derivative operators or more general field-dependent couplings. The
statement established here is narrower and more precise: in the minimal local
two-derivative system, the \(\partial F\) contribution is an independent
obstruction, and it disappears only at the symmetric endpoint.

%
%
%
%
%
%
%

%
%
%
%
%
%
%

%
%
%
%
%
%
%

%
%
%
%
%
%
%

\section{Spin--2 on curved backgrounds}
\label{sec:spin2_gravity_obstruction}

We now consider the bosonic analogue of the obstruction analysis carried out for
spin-\(\tfrac32\). The consistency of massive spin--2 propagation on curved
backgrounds has a substantial literature, especially in connection with
non-minimal curvature couplings and Einstein backgrounds
\cite{Buchbinder:1999ar,Buchbinder:1999gd,Deser:2006sq,Deser:2021qcg}. Our aim
in this section is therefore not to claim a new neutral spin--2 consistency
result in substance. Rather, we recast the neutral massive spin--2 problem in
the explicit language of the Fierz--Pauli subsidiary chain used throughout this
paper. This serves two purposes. First, it makes the comparison with the
spin-\(\tfrac32\) analysis completely transparent. Second, it prepares the next
step, namely the charged spin--2 system, where we will ask whether
electromagnetic couplings can produce a mechanism analogous to the
Einstein--Maxwell stress-tensor matching found for spin-\(\tfrac32\).

We start from the free massive Fierz--Pauli system for a symmetric rank-two
tensor in flat space and ask a basic question: is the system consistent if one
replaces ordinary derivatives by Levi--Civita covariant derivatives? The answer
is negative on a generic curved background. The failure appears already in the
divergence of the covariantized wave equation.

On a general background, the obstruction contains both curvature times first
derivatives of the field and derivatives of the Ricci tensor multiplying the
field itself. On Einstein spaces the derivative-of-Ricci terms disappear, but an
irreducible Weyl-coupled derivative term remains. We then show how this
Einstein-space Weyl obstruction is canceled within the standard local algebraic
class of curvature deformations. Finally, we analyze the trace and scalar
sectors and show that, for the deformed operator, the lower-spin subsidiary
system closes on Einstein backgrounds without any mass-curvature tuning. In
particular, there is no such residual Weyl obstruction on conformally flat
Einstein backgrounds, where the Weyl tensor vanishes identically.

We begin with the free massive Fierz--Pauli system in flat space for a symmetric
tensor \(\Phi_{\mu\nu}=\Phi_{\nu\mu}\):
\begin{equation}
(\partial^2+M^2)\Phi_{\mu\nu}=0,
\label{eq:spin2_flat_wave}
\end{equation}
supplemented by the subsidiary conditions
\begin{equation}
\partial^\mu \Phi_{\mu\nu}=0,
\qquad
\Phi^\mu{}_\mu=0.
\label{eq:spin2_flat_constraints}
\end{equation}
These equations remove the lower-spin components contained in a generic
symmetric tensor and propagate the correct five degrees of freedom of a massive
spin--2 particle.

\subsection{Naive gravitational covariantization}
\label{sec:naive_gravity_obstruction}

We now place the system on a curved background \((\mathcal M,g_{\mu\nu})\) with
torsionless Levi--Civita connection \(\nabla_\mu\), and consider the naive
covariantization
\begin{equation}
\partial_\mu \longrightarrow \nabla_\mu.
\end{equation}
This gives
\begin{equation}
E_{\mu\nu}\equiv (\nabla^2+M^2)\Phi_{\mu\nu}=0,
\label{eq:spin2_cov_wave}
\end{equation}
together with
\begin{equation}
J_\nu\equiv \nabla^\mu \Phi_{\mu\nu}=0,
\qquad
\Phi\equiv \Phi^\mu{}_\mu=0,
\label{eq:spin2_cov_constraints}
\end{equation}
where
\begin{equation}
\nabla^2\equiv \nabla^\rho\nabla_\rho.
\end{equation}
The trace condition remains algebraic. The nontrivial issue is whether the
divergence condition \(J_\nu=0\) is preserved by the evolution defined by
\eqref{eq:spin2_cov_wave}.

Taking the covariant divergence of \eqref{eq:spin2_cov_wave}, one finds
\begin{equation}
\nabla^\mu E_{\mu\nu}
=
\nabla^\mu (\nabla^2+M^2)\Phi_{\mu\nu}
=
(\nabla^2+M^2)J_\nu+[\nabla^\mu,\nabla^2]\Phi_{\mu\nu}.
\label{eq:spin2_div_basic}
\end{equation}
It is therefore natural to define
\begin{equation}
\mathcal O_\nu\equiv [\nabla^\mu,\nabla^2]\Phi_{\mu\nu},
\label{eq:spin2_obstruction_definition}
\end{equation}
so that
\begin{equation}
\nabla^\mu E_{\mu\nu}
=
(\nabla^2+M^2)J_\nu+\mathcal O_\nu.
\label{eq:spin2_divergence_with_obstruction}
\end{equation}
If \(\mathcal O_\nu\) could be expressed entirely in terms of the existing
subsidiary quantities \(J_\nu\), \(\Phi\), and their derivatives, then the
constraint chain would remain closed. The question is whether curvature
produces genuinely new structures.

For a symmetric rank-two tensor we use
\begin{equation}
[\nabla_\alpha,\nabla_\beta]\Phi_{\mu\nu}
=
R_{\mu\rho\alpha\beta}\,\Phi^\rho{}_\nu
+
R_{\nu\rho\alpha\beta}\,\Phi_\mu{}^\rho.
\label{eq:spin2_tensor_commutator}
\end{equation}
The operator commutator may be decomposed as
\begin{equation}
[\nabla^\mu,\nabla^2]
=
[\nabla^\mu,\nabla^\rho]\nabla_\rho
+
\nabla^\rho[\nabla^\mu,\nabla_\rho].
\label{eq:spin2_operator_split}
\end{equation}

From \eqref{eq:spin2_tensor_commutator} one obtains
\begin{equation}
[\nabla^\mu,\nabla_\rho]\Phi_{\mu\nu}
=
R_{\sigma\rho}\,\Phi^\sigma{}_\nu
+
R_{\nu\sigma\mu\rho}\,\Phi^{\mu\sigma}.
\label{eq:spin2_comm_phi}
\end{equation}
Taking one more derivative gives
\begin{align}
\nabla^\rho[\nabla^\mu,\nabla_\rho]\Phi_{\mu\nu}
&=
(\nabla^\rho R_{\sigma\rho})\Phi^\sigma{}_\nu
+
R_{\sigma\rho}\nabla^\rho\Phi^\sigma{}_\nu
\nonumber\\
&\qquad
+
(\nabla^\rho R_{\nu\sigma\mu\rho})\Phi^{\mu\sigma}
+
R_{\nu\sigma\mu\rho}\nabla^\rho\Phi^{\mu\sigma}.
\label{eq:spin2_second_piece}
\end{align}

For the first term in \eqref{eq:spin2_operator_split}, regard
\(\nabla_\rho\Phi_{\mu\nu}\) as a rank-three covariant tensor and apply the
commutator to each of its tensor indices. Using the symmetry
\(\Phi_{\mu\nu}=\Phi_{\nu\mu}\), one finds
\begin{equation}
[\nabla^\mu,\nabla^\rho]\nabla_\rho\Phi_{\mu\nu}
=
2R^{\sigma\rho}\nabla_\rho\Phi_{\sigma\nu}
+
R_{\nu\sigma\mu\rho}\nabla^\rho\Phi^{\mu\sigma}.
\label{eq:spin2_first_piece}
\end{equation}

Combining \eqref{eq:spin2_second_piece} and
\eqref{eq:spin2_first_piece}, one obtains the exact obstruction
\begin{equation}
\mathcal O_\nu
=
3R^{\sigma\rho}\nabla_\rho\Phi_{\sigma\nu}
+
2R_{\nu\sigma\mu\rho}\nabla^\rho\Phi^{\mu\sigma}
+
(\nabla^\rho R_{\sigma\rho})\Phi^\sigma{}_\nu
+
(\nabla^\rho R_{\nu\sigma\mu\rho})\Phi^{\mu\sigma}.
\label{eq:spin2_obstruction_exact}
\end{equation}
This already shows that the divergence of the naively covariantized wave
equation is not homogeneous in the subsidiary quantity \(J_\nu\). In addition to
curvature times \(\nabla\Phi\), one generates derivative-of-curvature terms.

The last term may be rewritten with the contracted Bianchi identity,
\begin{equation}
\nabla^\rho R_{\nu\sigma\mu\rho}
=
\nabla_\sigma R_{\mu\nu}
-
\nabla_\nu R_{\mu\sigma}.
\label{eq:spin2_contracted_bianchi}
\end{equation}
Therefore
\begin{equation}
(\nabla^\rho R_{\nu\sigma\mu\rho})\Phi^{\mu\sigma}
=
(\nabla_\sigma R_{\mu\nu}-\nabla_\nu R_{\mu\sigma})\Phi^{\mu\sigma}.
\label{eq:spin2_ricci_derivative_term}
\end{equation}
Because \(\Phi^{\mu\sigma}\) is symmetric, this does not vanish identically. The
second term carries the free index \(\nu\), so the two contributions are not
related by a dummy-index relabeling.

It follows that on a generic curved background the divergence equation contains
two independent curvature-generated structures,
\begin{equation}
(Riemann)\cdot \nabla\Phi,
\qquad
(\nabla Ricci)\cdot \Phi.
\label{eq:spin2_schematic_generic}
\end{equation}
These are not proportional to the original Fierz--Pauli subsidiary quantities.
Hence the naively covariantized Fierz--Pauli system does not close in general.

The obstruction simplifies on Einstein spaces,
\begin{equation}
R_{\mu\nu}=\Lambda g_{\mu\nu},
\qquad
\nabla_\rho R_{\mu\nu}=0.
\label{eq:spin2_einstein_bg}
\end{equation}
In that case the derivative-of-Ricci terms vanish, and the Ricci term in
\eqref{eq:spin2_obstruction_exact} becomes
\begin{equation}
3R^{\sigma\rho}\nabla_\rho\Phi_{\sigma\nu}=3\Lambda J_\nu.
\end{equation}
Therefore
\begin{equation}
\mathcal O_\nu
=
3\Lambda J_\nu
+
2R_{\nu\sigma\mu\rho}\nabla^\rho\Phi^{\mu\sigma}.
\label{eq:spin2_obstruction_einstein_1}
\end{equation}

We now decompose the Riemann tensor into Weyl and constant-curvature parts. In
four dimensions, on an Einstein background,
\begin{equation}
R_{\mu\nu\rho\sigma}
=
C_{\mu\nu\rho\sigma}
+
\frac{\Lambda}{3}
\left(
g_{\mu\rho}g_{\nu\sigma}-g_{\mu\sigma}g_{\nu\rho}
\right),
\label{eq:spin2_riemann_einstein}
\end{equation}
where \(C_{\mu\nu\rho\sigma}\) is the Weyl tensor. Substituting
\eqref{eq:spin2_riemann_einstein} into
\eqref{eq:spin2_obstruction_einstein_1}, one obtains
\begin{align}
\mathcal O_\nu
&=
3\Lambda J_\nu
+
2C_{\nu\sigma\mu\rho}\nabla^\rho\Phi^{\mu\sigma}
\nonumber\\
&\qquad
+
\frac{2\Lambda}{3}
\left(
g_{\nu\mu}g_{\sigma\rho}-g_{\nu\rho}g_{\sigma\mu}
\right)\nabla^\rho\Phi^{\mu\sigma}
\nonumber\\
&=
3\Lambda J_\nu
+
2C_{\nu\sigma\mu\rho}\nabla^\rho\Phi^{\mu\sigma}
+
\frac{2\Lambda}{3}\left(J_\nu-\nabla_\nu\Phi\right).
\label{eq:spin2_obstruction_einstein_2}
\end{align}
Using the trace constraint \(\Phi=0\), this reduces to
\begin{equation}
\mathcal O_\nu
=
\frac{11\Lambda}{3}J_\nu
+
2C_{\nu\sigma\mu\rho}\nabla^\rho\Phi^{\mu\sigma}.
\label{eq:spin2_obstruction_einstein_final}
\end{equation}
The divergence equation therefore takes the form
\begin{equation}
\nabla^\mu E_{\mu\nu}
=
\left(\nabla^2+M^2+\frac{11\Lambda}{3}\right)J_\nu
+
2C_{\nu\sigma\mu\rho}\nabla^\rho\Phi^{\mu\sigma}.
\label{eq:spin2_divergence_einstein_final}
\end{equation}
Once the trace and divergence constraints are imposed, the remaining obstruction
is the Weyl-coupled derivative term
\begin{equation}
\mathcal O_\nu\big|_{J=0=\Phi}
=
2C_{\nu\sigma\mu\rho}\nabla^\rho\Phi^{\mu\sigma}.
\label{eq:spin2_weyl_obstruction}
\end{equation}
Thus naive gravitational covariantization fails even on Einstein backgrounds,
unless the background is maximally symmetric. In particular, there is no such
residual obstruction on conformally flat Einstein backgrounds, for which
\(C_{\mu\nu\rho\sigma}=0\).

This is the form in which the neutral spin--2 result is useful for the next
section. In the subsidiary-chain language, Einstein space removes the
Ricci-derivative obstruction, while the remaining obstruction is a purely
Weyl-controlled differential term. The charged case will then allow us to ask
whether the electromagnetic sector changes this structure in any essential way.

\subsection{Most general local curvature deformation and the Weyl obstruction}
\label{subsec:spin2_curvature_deformation}

We now ask whether this Einstein-space Weyl obstruction can be removed by local
algebraic curvature couplings. Restricting to terms with no derivatives acting
on \(\Phi_{\mu\nu}\) and at most one power of the background curvature, the most
general symmetric rank-two deformation is
\begin{align}
\Delta E_{\mu\nu}
&=
a\,R_{\mu\rho\nu\sigma}\Phi^{\rho\sigma}
+b\,R_{\rho(\mu}\Phi_{\nu)}{}^\rho
+c\,R\,\Phi_{\mu\nu}
+d\,g_{\mu\nu}R_{\rho\sigma}\Phi^{\rho\sigma}
+e\,g_{\mu\nu}R\,\Phi
+f\,R_{\mu\nu}\Phi .
\label{eq:spin2_general_curv_def}
\end{align}
The deformed equation is
\begin{equation}
\widehat E_{\mu\nu}\equiv E_{\mu\nu}+\Delta E_{\mu\nu}=0.
\label{eq:spin2_deformed_operator}
\end{equation}

Taking the divergence of \eqref{eq:spin2_general_curv_def} gives
\begin{align}
\nabla^\mu \Delta E_{\mu\nu}
&=
a\,(\nabla^\mu R_{\mu\rho\nu\sigma})\Phi^{\rho\sigma}
+a\,R_{\mu\rho\nu\sigma}\nabla^\mu\Phi^{\rho\sigma}
\nonumber\\
&\qquad
+b\,\nabla^\mu\!\left(R_{\rho(\mu}\Phi_{\nu)}{}^\rho\right)
+c\,\nabla^\mu(R\Phi_{\mu\nu})
+d\,\nabla_\nu(R_{\rho\sigma}\Phi^{\rho\sigma})
+e\,\nabla_\nu(R\Phi)
+f\,\nabla^\mu(R_{\mu\nu}\Phi).
\label{eq:spin2_div_delta_general}
\end{align}
On a generic background this still contains derivative-of-curvature terms. Thus
a purely algebraic deformation does not solve the generic problem; it only
reshuffles it.

On Einstein spaces,
\begin{equation}
R_{\mu\nu}=\Lambda g_{\mu\nu},
\qquad
R=4\Lambda,
\qquad
\nabla_\rho R_{\mu\nu}=0,
\qquad
\nabla_\rho R=0,
\label{eq:spin2_einstein_again}
\end{equation}
the deformation simplifies to
\begin{equation}
\nabla^\mu \Delta E_{\mu\nu}
=
a\,R_{\mu\rho\nu\sigma}\nabla^\mu\Phi^{\rho\sigma}
+
(b+4c)\Lambda\,J_\nu
+
(d+4e+f)\Lambda\,\nabla_\nu\Phi .
\label{eq:spin2_div_delta_einstein_1}
\end{equation}
Thus only the Riemann coupling can affect the Weyl sector. The remaining terms
contribute only to the \(J_\nu\) and \(\nabla_\nu\Phi\) channels.

Using again the Einstein decomposition
\eqref{eq:spin2_riemann_einstein}, one finds
\begin{align}
a\,R_{\mu\rho\nu\sigma}\nabla^\mu\Phi^{\rho\sigma}
&=
a\,C_{\mu\rho\nu\sigma}\nabla^\mu\Phi^{\rho\sigma}
+\frac{a\Lambda}{3}
\left(
g_{\mu\nu}g_{\rho\sigma}-g_{\mu\sigma}g_{\rho\nu}
\right)\nabla^\mu\Phi^{\rho\sigma}
\nonumber\\
&=
-a\,C_{\nu\sigma\mu\rho}\nabla^\rho\Phi^{\mu\sigma}
+\frac{a\Lambda}{3}\left(\nabla_\nu\Phi-J_\nu\right),
\label{eq:spin2_a_term_einstein}
\end{align}
where we used the algebraic symmetries of the Weyl tensor in the first term.

Substituting \eqref{eq:spin2_a_term_einstein} into
\eqref{eq:spin2_div_delta_einstein_1}, one obtains
\begin{equation}
\nabla^\mu \Delta E_{\mu\nu}
=
-a\,C_{\nu\sigma\mu\rho}\nabla^\rho\Phi^{\mu\sigma}
+
\left(-\frac{a}{3}+b+4c\right)\Lambda\,J_\nu
+
\left(\frac{a}{3}+d+4e+f\right)\Lambda\,\nabla_\nu\Phi .
\label{eq:spin2_div_delta_einstein_final}
\end{equation}
Adding this to \eqref{eq:spin2_divergence_einstein_final} gives
\begin{align}
\nabla^\mu \widehat E_{\mu\nu}
&=
\left[
\nabla^2+M^2
+\left(
\frac{11}{3}-\frac{a}{3}+b+4c
\right)\Lambda
\right]J_\nu
\nonumber\\
&\qquad
+
\left(
-\frac{2}{3}+\frac{a}{3}+d+4e+f
\right)\Lambda\,\nabla_\nu\Phi
+
(2-a)\,C_{\nu\sigma\mu\rho}\nabla^\rho\Phi^{\mu\sigma}.
\label{eq:spin2_div_deformed_einstein}
\end{align}
The Weyl-coupled derivative term is therefore canceled if and only if
\begin{equation}
a=2.
\label{eq:spin2_unique_a}
\end{equation}
Within the local algebraic class \eqref{eq:spin2_general_curv_def}, no Ricci or
scalar-curvature deformation contributes to the Weyl sector on an Einstein
background.

With the choice \(a=2\), the divergence equation becomes
\begin{equation}
\nabla^\mu \widehat E_{\mu\nu}
=
\left[
\nabla^2+M^2+(3+b+4c)\Lambda
\right]J_\nu
+
\left(d+4e+f\right)\Lambda\,\nabla_\nu\Phi .
\label{eq:spin2_div_deformed_einstein_a2}
\end{equation}
To make the vector subsidiary equation homogeneous, one must impose
\begin{equation}
d+4e+f=0.
\label{eq:spin2_D_zero}
\end{equation}
Then
\begin{equation}
\nabla^\mu \widehat E_{\mu\nu}
=
\left[
\nabla^2+M^2+(3+b+4c)\Lambda
\right]J_\nu .
\label{eq:spin2_divergence_closed}
\end{equation}
Thus the divergence constraint propagates homogeneously.

We now analyze the trace and scalar sectors. On an Einstein background, the
operator itself reduces to
\begin{equation}
\widehat E_{\mu\nu}
=
\left[
\nabla^2+M^2+\left(b+4c-\frac23\right)\Lambda
\right]\Phi_{\mu\nu}
+
2\,C_{\mu\rho\nu\sigma}\Phi^{\rho\sigma}
+
\left(d+4e+f+\frac23\right)\Lambda\,g_{\mu\nu}\Phi.
\label{eq:spin2_operator_reduced}
\end{equation}
Taking the trace and using the tracelessness of the Weyl tensor gives
\begin{equation}
g^{\mu\nu}\widehat E_{\mu\nu}
=
\left[
\nabla^2+M^2+\left(2+b+4c\right)\Lambda
\right]\Phi,
\label{eq:spin2_trace_closed}
\end{equation}
where we also used \eqref{eq:spin2_D_zero}. Thus the trace condition is also
propagated homogeneously.

To examine the next scalar step, take one more divergence of
\eqref{eq:spin2_divergence_closed}:
\begin{equation}
\nabla^\nu\nabla^\mu \widehat E_{\mu\nu}
=
\nabla^\nu
\left(
\left[
\nabla^2+M^2+(3+b+4c)\Lambda
\right]J_\nu
\right).
\label{eq:spin2_double_div_start}
\end{equation}
Since \(M\), \(\Lambda\), \(b\), and \(c\) are constants on the Einstein
background, this becomes
\begin{equation}
\nabla^\nu\nabla^\mu \widehat E_{\mu\nu}
=
\left[
\nabla^2+M^2+(3+b+4c)\Lambda
\right]\nabla^\nu J_\nu
+
[\nabla^\nu,\nabla^2]J_\nu.
\label{eq:spin2_double_div_comm}
\end{equation}
For a covector \(J_\nu\) on an Einstein background one has
\begin{equation}
[\nabla^\nu,\nabla^2]J_\nu=\Lambda\,\nabla^\nu J_\nu.
\label{eq:spin2_vector_commutator}
\end{equation}
Therefore
\begin{equation}
\nabla^\nu\nabla^\mu \widehat E_{\mu\nu}
=
\left[
\nabla^2+M^2+(4+b+4c)\Lambda
\right]\nabla^\nu J_\nu .
\label{eq:spin2_double_divergence_closed}
\end{equation}
This is again homogeneous in the next lower-spin scalar quantity
\(\nabla^\nu J_\nu\). No new source term proportional to \(\Phi\), and no new
curvature obstruction, appears at this stage.

We may now summarize the result. On a generic curved background, the naively
covariantized spin--2 system does not close. On Einstein spaces, the
derivative-of-Ricci obstruction disappears, but a Weyl-coupled derivative term
remains. For conformally flat Einstein backgrounds this residual obstruction is
absent from the start. More generally, within the standard local algebraic class
of curvature deformations, the Weyl term is canceled by the Riemann coupling
with coefficient
\begin{equation}
a=2,
\end{equation}
while homogeneous propagation of the vector subsidiary equation further requires
\begin{equation}
d+4e+f=0.
\end{equation}
With these conditions, the deformed operator has a homogeneous divergence
equation, a homogeneous trace equation, and a homogeneous double-divergence
equation:
\begin{align}
\nabla^\mu \widehat E_{\mu\nu}
&=
\left[
\nabla^2+M^2+(3+b+4c)\Lambda
\right]J_\nu,
\\[1mm]
g^{\mu\nu}\widehat E_{\mu\nu}
&=
\left[
\nabla^2+M^2+(2+b+4c)\Lambda
\right]\Phi,
\\[1mm]
\nabla^\nu\nabla^\mu \widehat E_{\mu\nu}
&=
\left[
\nabla^2+M^2+(4+b+4c)\Lambda
\right]\nabla^\nu J_\nu.
\end{align}
Thus the lower-spin subsidiary system associated with \(\widehat E_{\mu\nu}\)
closes on Einstein backgrounds for
\begin{equation}
a=2,
\qquad
d+4e+f=0.
\label{eq:spin2_final_closure_conditions}
\end{equation}
The combination \(b+4c\) remains free and only shifts the effective
curvature-dependent masses in the homogeneous propagation equations for
\(J_\nu\), \(\Phi\), and \(\nabla^\nu J_\nu\). In particular, unlike the
spin-\(\tfrac32\) Einstein--Maxwell problem, no mass-curvature tuning is required
for closure.

This completes the neutral bosonic comparison needed here and prepares the
charged extension of the next section. For neutral spin--2, the relevant curved
background obstruction is differential and Weyl-controlled; it never reorganizes
into an algebraic rank-two channel. This is the structural contrast we will need
when asking whether electromagnetism can change the spin--2 story in a way
analogous to the spin-\(\tfrac32\) Einstein--Maxwell mechanism.

\section{Charged spin--2 on curved backgrounds}
\label{sec:spin2_charged_curved}

We now extend the bosonic analysis to the case of a charged massive spin--2
field in a curved spacetime. Charged spin--2 fields on non-flat backgrounds have
been discussed previously, in particular in effective-field-theory constructions
on fixed Einstein backgrounds with non-minimal curvature and electromagnetic
couplings; see for example appendix A.2 of \cite{Benini:2010pr}. Those analyses
start from a chosen Lagrangian and study degree-of-freedom counting, typically
within a weak-field regime. Our purpose here is different. We analyze the system
directly in the Fierz--Pauli subsidiary-chain language used throughout this
paper, and ask two questions.

First, does there exist an exact charged massive spin--2 Fierz--Pauli system on a
curved background whose lower-spin chain closes? Second, if such a system exists,
does its closure involve any analogue of the spin-\(\tfrac32\)
Einstein--Maxwell stress-tensor matching mechanism?

We will find that the answer to the first question is yes, under the same
Einstein-space curvature repair as in the neutral case and with the charged flat-space
gyromagnetic condition \(\gamma=1\). The answer to the second question is no:
although the charged spin--2 chain closes, its obstruction remains differential
and never reorganizes into an algebraic Einstein--Maxwell tensor channel.
Consequently, no tuned locus relating mass, charge, and curvature emerges.

We stress that this obstruction is not a specifically charged phenomenon. It is
already present for the neutral spin--2 field under naive gravitational
covariantization. The role of charge is therefore not the main issue here. The
point is instead that for a rank-two bosonic field the gravitational
obstruction appears first in a differential channel, through curvature times
\(\nabla\Phi\) and, on generic backgrounds, through derivatives of the Ricci
tensor. On Einstein spaces this reduces to the Weyl-coupled derivative term,
which vanishes only in special limits, such as maximally symmetric backgrounds
and, more generally, conformally flat Einstein backgrounds.

\subsection{Charged Fierz--Pauli system in flat spacetime}

Let \(\Phi_{\mu\nu}=\Phi_{\nu\mu}\) be a complex symmetric tensor. The charged
Fierz--Pauli system in a constant electromagnetic background is
\begin{equation}
E_{\mu\nu}
\equiv
(D^2+M^2)\Phi_{\mu\nu}
-2 i e\,\gamma\,
\bigl(
F_{\mu}{}^{\rho}\Phi_{\rho\nu}
-
\Phi_{\mu}{}^{\rho}F_{\rho\nu}
\bigr)
=0,
\label{eq:spin2_charged_flat}
\end{equation}
supplemented by the subsidiary conditions
\begin{equation}
J_\nu \equiv D^\mu\Phi_{\mu\nu}=0,
\qquad
\Phi \equiv \Phi^\mu{}_\mu=0.
\label{eq:spin2_charged_constraints}
\end{equation}

Taking the divergence of \eqref{eq:spin2_charged_flat}, one finds
\begin{equation}
D^\mu E_{\mu\nu}
=
(D^2+M^2)J_\nu
+
2 i e (1-\gamma)\,F^\mu{}_{\rho}D_\mu\Phi^\rho{}_\nu
-
2 i e \gamma\,F_{\nu\rho}J^\rho.
\label{eq:spin2_flat_divergence_charged}
\end{equation}
The term \(F^\mu{}_{\rho}D_\mu\Phi^\rho{}_\nu\) is independent of \(J_\nu\).
Closure of the flat-space subsidiary system therefore requires
\begin{equation}
\gamma=1,
\label{eq:spin2_gamma_one}
\end{equation}
which corresponds to gyromagnetic ratio \(g=2\). For \(\gamma=1\), the flat-space
divergence becomes homogeneous,
\begin{equation}
D^\mu E_{\mu\nu}
=
(D^2+M^2)J_\nu
-
2 i e\,F_{\nu\rho}J^\rho .
\label{eq:spin2_flat_divergence_closed}
\end{equation}

\subsection{Exact closure on Einstein backgrounds with covariantly constant \texorpdfstring{$F$}{F}}

We now place the system on a curved background and replace
\begin{equation}
D_\mu=\nabla_\mu+i e A_\mu.
\end{equation}
As in the neutral case, we allow the most general local algebraic curvature
deformation with at most one power of the background curvature,
\begin{align}
\Delta E_{\mu\nu}
&=
a\,R_{\mu\rho\nu\sigma}\Phi^{\rho\sigma}
+b\,R_{\rho(\mu}\Phi_{\nu)}{}^\rho
+c\,R\,\Phi_{\mu\nu}
+d\,g_{\mu\nu}R_{\rho\sigma}\Phi^{\rho\sigma}
+e\,g_{\mu\nu}R\,\Phi
+f\,R_{\mu\nu}\Phi .
\label{eq:spin2_charged_general_curv_def}
\end{align}
The deformed charged operator is
\begin{equation}
\widehat E_{\mu\nu}
=
(D^2+M^2)\Phi_{\mu\nu}
-2 i e\,\gamma\,
\bigl(
F_{\mu}{}^{\rho}\Phi_{\rho\nu}
-
\Phi_{\mu}{}^{\rho}F_{\rho\nu}
\bigr)
+\Delta E_{\mu\nu}.
\label{eq:spin2_charged_deformed_operator}
\end{equation}

We restrict to backgrounds with covariantly constant field strength,
\begin{equation}
\nabla_\rho F_{\mu\nu}=0,
\label{eq:spin2_cov_constF}
\end{equation}
and to Einstein spaces,
\begin{equation}
R_{\mu\nu}=\Lambda g_{\mu\nu}.
\label{eq:spin2_charged_einstein_bg}
\end{equation}
Under these assumptions, the divergence of the deformed operator combines the
neutral Einstein-space result of section~\ref{sec:spin2_gravity_obstruction}
with the flat charged contribution and gives
\begin{align}
D^\mu \widehat E_{\mu\nu}
&=
\left[
D^2+M^2+
\left(
\frac{11}{3}-\frac{a}{3}+b+4c
\right)\Lambda
\right]J_\nu
\nonumber\\
&\qquad
+
\left(
-\frac{2}{3}+\frac{a}{3}+d+4e+f
\right)\Lambda\,D_\nu\Phi
+
(2-a)\,C_{\nu\sigma\mu\rho}D^\rho\Phi^{\mu\sigma}
\nonumber\\
&\qquad
+
2 i e (1-\gamma)\,F^\mu{}_{\rho}D_\mu\Phi^\rho{}_\nu
-
2 i e \gamma\,F_{\nu\rho}J^\rho .
\label{eq:spin2_charged_curved_divergence_general}
\end{align}

This formula makes the structure transparent. There are three independent ways in
which closure can fail:
\begin{equation}
D_\nu\Phi,
\qquad
C_{\nu\sigma\mu\rho}D^\rho\Phi^{\mu\sigma},
\qquad
F^\mu{}_{\rho}D_\mu\Phi^\rho{}_\nu.
\label{eq:spin2_charged_three_obstructions}
\end{equation}

The first two are the neutral spin--2 obstructions already discussed in
section~\ref{sec:spin2_gravity_obstruction}; the third is the charged flat-space
obstruction. Exact closure of the vector subsidiary equation therefore requires
\begin{equation}
\gamma=1,
\qquad
a=2,
\qquad
d+4e+f=0.
\label{eq:spin2_charged_closure_conditions}
\end{equation}
With these conditions imposed, the divergence reduces to the homogeneous equation
\begin{equation}
D^\mu \widehat E_{\mu\nu}
=
\left[
D^2+M^2+(3+b+4c)\Lambda
\right]J_\nu
-
2 i e\,F_{\nu\rho}J^\rho .
\label{eq:spin2_charged_divergence_closed}
\end{equation}
Thus the vector subsidiary condition \(J_\nu=0\) propagates consistently.

The trace is even simpler. Since the Pauli term is antisymmetric in the vector
indices, it drops out identically from the trace. One finds
\begin{equation}
g^{\mu\nu}\widehat E_{\mu\nu}
=
\left[
D^2+M^2+(2+b+4c)\Lambda
\right]\Phi ,
\label{eq:spin2_charged_trace_closed}
\end{equation}
where we have again used \(d+4e+f=0\). Thus the trace constraint \(\Phi=0\) also
propagates homogeneously.

At this point the lower-spin chain is already closed: both subsidiary quantities
\(J_\nu\) and \(\Phi\) satisfy homogeneous propagation equations. Any further
divergence of \eqref{eq:spin2_charged_divergence_closed} is therefore
automatically homogeneous in \(J_\nu\), \(\Phi\), and their derivatives, and no
new independent lower-spin obstruction is generated.

We may summarize the result as follows. On Einstein backgrounds with
covariantly constant electromagnetic field strength, the charged massive spin--2
Fierz--Pauli system closes exactly provided
\begin{equation}
\gamma=1,
\qquad
a=2,
\qquad
d+4e+f=0.
\label{eq:spin2_charged_exact_closure}
\end{equation}
In particular, unlike the spin-\(\tfrac32\) Einstein--Maxwell case, no relation
among mass, charge, and curvature is required for closure.

\subsection{Absence of Einstein--Maxwell stress-tensor matching}

The essential difference between  spin--2 Fierz--Pauli chain and the the spin-\(\tfrac32\) one is seen directly 
in the form of the divergence
\eqref{eq:spin2_charged_curved_divergence_general}. The charged sector modifies
the subsidiary equation only through derivative terms such as
\begin{equation}
F^\mu{}_{\rho}D_\mu\Phi^\rho{}_\nu,
\label{eq:spin2_charged_derivative_obstruction}
\end{equation}
and, after repair, through the homogeneous mixing term
\begin{equation}
F_{\nu\rho}J^\rho .
\label{eq:spin2_charged_homogeneous_mixing}
\end{equation}
It never produces an algebraic rank-two tensor of the form \(T^{(F)}_{\mu\nu}\Phi^\nu\).

One can ask whether such a channel might arise from additional local EFT
corrections. At quadratic order in the field strength, the most general
algebraic operators built from \(F_{\mu\nu}\) and \(\Phi_{\mu\nu}\) include
terms such as
\begin{equation}
\Delta E_{\mu\nu}^{(F^2)}
=
\alpha_1\,F_{\mu\rho}F_\nu{}^\rho\,\Phi
+
\alpha_2\,g_{\mu\nu}F^2\Phi
+
\alpha_3\,F_{\mu\rho}F_{\nu\sigma}\Phi^{\rho\sigma},
\label{eq:spin2_F2_operators}
\end{equation}
and analogous higher-derivative completions. However, taking a divergence and
using \(\nabla F=0\), one always obtains terms with at least one derivative on
the field, schematically
\begin{equation}
F^2\,D\Phi,
\label{eq:spin2_F2_divergence_structure}
\end{equation}
or higher-derivative analogues. Even if one uses the equations of motion to
trade \(D^2\Phi\sim M^2\Phi+\cdots\), one derivative necessarily remains.

This reflects a general derivative-counting property of second-order systems:
for a local operator whose leading kinetic term is \(D^2\Phi\), the divergence
\(D^\mu E_{\mu\nu}\) necessarily contains at least one derivative acting on
\(\Phi\), modulo terms proportional to the equations of motion. Consequently, no
algebraic rank-two tensor of the form
\begin{equation}
\mathcal T_{\mu\nu}\Phi^\nu,
\qquad
\mathcal T_{\mu\nu}\sim T^{(F)}_{\mu\nu},
\label{eq:spin2_rank_two_channel}
\end{equation}
can be generated in the divergence. In particular, although
\begin{equation}
F_{\mu\rho}F_\nu{}^\rho
=
T^{(F)}_{\mu\nu}
+\frac14 g_{\mu\nu}F^2
\label{eq:spin2_TF_decomposition}
\end{equation}
appears at the operator level, it never enters the divergence in a way that
would allow matching with the gravitational sector.

Thus the charged spin--2 system closes, but its closure mechanism is
qualitatively different from that of spin-\(\tfrac32\): there is no algebraic
Einstein--Maxwell stress-tensor channel, and therefore no tuned locus.

\subsection{Comparison with spin-\texorpdfstring{$\tfrac32$}{3/2} and connection to the \texorpdfstring{$\beta_\kappa$}{beta-kappa} framework}

The origin of this difference can be traced to the order of the kinetic
operator.

For spin-\(\tfrac32\), the equations are first order. The divergence therefore
produces terms schematically of the form
\begin{equation}
D\mathcal R \sim D^2\Psi + F\,D\Psi.
\end{equation}
Using the first-order equations of motion, the differential term can be reduced,
and the algebra closes on an algebraic rank-two contribution,
\begin{equation}
F^2\Psi \sim T^{(F)}_{\mu\nu}\Psi^\nu.
\end{equation}
This is precisely what allows matching with the Einstein-tensor channel and
leads to the tuned locus.

By contrast, for spin--2 the operator is second order, and the divergence always
retains at least one derivative acting on the field. The charged system can be
repaired so that its Fierz--Pauli chain closes, but the obstruction remains
intrinsically differential and never reorganizes into an algebraic
Einstein--Maxwell tensor channel. In the language of the \(\beta_\kappa\)
framework introduced for spin-\(\tfrac32\), the bosonic system lacks the
crucial ingredient that allowed interpolation between decoupled and
supergravity regimes: a Pauli sector capable of reproducing the same tensor
structure as the gravitational channel.

More concretely:
\begin{itemize}
\item In the spin-\(\tfrac32\) case, the parameter \(\beta_\kappa\) controls the
interpolation between a decoupled system and a supergravity-completed system,
with closure selecting a specific locus through stress-tensor matching.

\item In the spin--2 case, the charged curved-space Fierz--Pauli chain closes
already for
\begin{equation}
\gamma=1,
\qquad
a=2,
\qquad
d+4e+f=0,
\end{equation}
with no relation between mass, charge, and curvature. The electromagnetic and
gravitational sectors are repaired independently, not by matching a common
algebraic tensor channel.
\end{itemize}

We may therefore summarize as follows. The absence of a
\(\beta_\kappa\)-type interpolation in the spin--2 system is a direct
consequence of the second-order nature of its kinetic operator. Unlike the
spin-\(\tfrac32\) case, the charged spin--2 subsidiary chain can close without
producing any algebraic Einstein--Maxwell matching channel, and no tuned locus
emerges.

\section{Conclusions}
\label{sec:conclusions}

We started this paper with two protagonists: the \(\cN=2\) supergravity
gravitino and the decoupled system derived from the first massive open-string
modes in a constant electromagnetic background. Both describe a charged massive
spin-$\tfrac32$ field, but they organize its lower-spin sector in very different
ways. The question was not whether the special supergravity mass--charge
relation is known --- it is --- but why it reappears when one studies the
covariant Fierz--Pauli chain of an isolated charged spin-$\tfrac32$ field on
curved backgrounds, and what this tells us about the range of validity of the
decoupled description.

A first lesson is that, once gravity is allowed to react, mass and charge are no
longer arbitrary. In the minimal Einstein--Maxwell setting studied here, the
consistency of the lower-spin chain forces the tuned relation
\begin{equation}
m^2+\frac{\Lambda}{3}=\frac{2e^2}{\kappa^2}.
\end{equation}
At \(\Lambda=0\), this places the mass at the Planck scale for a charge of order
the elementary electric charge, up to convention-dependent normalizations of the
gauge coupling. This locus has a long history. It already appears in early
supergravity analyses of charged spin-$\tfrac32$ couplings
\cite{Das:1976ct,Deser:1977uq}, was later identified as the causal point for
propagation on Einstein--Maxwell backgrounds \cite{Deser:2001dt}, and has more
recently reappeared in on-shell and positivity-based analyses of isolated
spin-$\tfrac32$ effective field theories
\cite{Gherghetta:2025tlx,Bellazzini:2025shd}. What is new here is not the
existence of the locus itself, but the covariant mechanism by which it is
enforced inside the Fierz--Pauli constraint chain.

The present analysis is covariant, off shell, and performed directly on curved
Einstein--Maxwell($-\Lambda$) backgrounds. The essential point is that
curvature commutators generate, inside the secondary constraint diagnostic, a
Ricci-built rank-two channel which, after using Einstein's equation, becomes
the Maxwell stress tensor. Closure then requires the Pauli sector to produce
the same traceless symmetric tensor, carried in four dimensions uniquely by the
mixed self-dual/anti-self-dual contraction. This explains why the maximally
asymmetric organization of the decoupled system cannot survive once gravity
reacts: it does not have access to the channel needed to match the
gravitational obstruction. In this sense, the supergravity value is recovered
here as a Fierz--Pauli consistency condition. The resemblance of the tuned locus
to a BPS relation is immediate, but here it is obtained purely from closure of
the covariant Fierz--Pauli chain, without assuming preserved supersymmetry.

The interpolating \(\beta_\kappa\)-family makes this structure especially
transparent. The Einstein--Maxwell stress-tensor channel can be matched by a
continuous family of systems, but the exact constant-field second-order
analysis shows that this is not sufficient. Although the entire family
reproduces \(g=2\) at linear order, a residual chiral \(F^2\) term is
generically present at quadratic order. Its coefficient vanishes only at the
two endpoints, \(\beta_\kappa=0\) and \(\beta_\kappa=1\). The first is, by
construction, the decoupled-gravity endpoint; the second is the symmetric
supergravity one. Once dynamical gravity is required, only the latter remains.
The analysis of non-constant electromagnetic backgrounds reaches the same
conclusion from a different direction: only the \(\beta_\kappa=1\) endpoint
avoids the additional \(\partial F\) obstruction. The supergravity
organization is therefore not selected by a single argument, but by the
combined effect of stress-tensor matching, exact quadratic structure, and
non-constant-field closure.

The flat-space tuned relation also has an immediate physical meaning. If the
charge is of order unity, for instance of the order of the usual electric
charge, then the tuned mass is Planckian. This is not useful for describing
ordinary elementary charged spin-$\tfrac32$ particles, where such a mass would
be phenomenologically unacceptable. A much lighter particle would require a
very small coupling, which may be plausible for a graviphoton but not for the
original problem that motivated the decoupled system. This point should not be
overstated, however. The present result does not imply that every charged
spin-$\tfrac32$ particle must have a Planckian mass. It states only that,
within the minimal framework studied here --- an isolated local two-derivative
spin-$\tfrac32$ system coupled only to Einstein--Maxwell gravity with constant
Fierz--Pauli data --- the consistency locus lies precisely where gravitational
backreaction becomes strong. The Schwarzschild radius is then of the same order
as the Compton wavelength. In that sense, the tuned relation is not a useful
low-energy mass formula for elementary charged matter; it is a consistency
condition for the coupled gravitational system.

This brings us back to the second protagonist. In flat-space particle physics,
where gravity is negligible, the decoupled Fierz--Pauli system remains the
appropriate description, and there is no reason to expect a Planck-suppressed
gravitational consistency condition to constrain it. What we have learned is
instead that its domain of applicability is narrower than its flat-space form
suggests. Even before one reaches the scale where standard four-point unitarity
arguments become relevant, its lower-spin organization must already be modified
if one insists on coupling it minimally to gravity. A consistent extension
therefore requires enlarging the framework beyond an isolated local
constant-parameter two-derivative Fierz--Pauli system.

This is exactly what one expects in string theory. The decoupled constant-field
system arises in the open-string limit, where the string scale and the Planck
scale can be parametrically separated. But the consistent parent theory contains
many additional degrees of freedom and, in the full theory, an infinite tower of
massive higher-spin states. In the language used here, these additional states
and interactions provide extra channels that can modify the effective
constraint algebra before the isolated local spin-$\tfrac32$ description fails.
Our analysis is fully compatible with that picture: it isolates the geometric
channel that must be matched once gravity reacts, but it does not exclude
completions in which other states participate in the matching.

The same remark applies to composite spin-$\tfrac32$ states. Massive charged
spin-$\tfrac32$ particles do exist in nature as hadronic resonances, such as the
\(\Omega^-\). Their effective description is not governed by a fundamental
\(g=2\) condition, nor by the supergravity tuned locus. Their completion is
QCD, which provides strong dynamics, additional degrees of freedom, and
interactions. This is exactly what one should expect: such states are not
isolated weakly coupled elementary particles with a large separation of scales.
At the same time, the decoupled system remains a natural starting point for
organizing their low-energy effective description~\cite{Benakli:2025xao}.

The purely gravitational limit is also instructive. Setting \(F_{\mu\nu}=0\),
the spin-$\tfrac32$ chain reduces to the requirement that the background be
Einstein. This reproduces the same substance as Buchdahl's classical
consistency condition, but in a more systematic organization: here it appears
directly as closure of the covariant Fierz--Pauli chain, rather than as
compatibility of the naively covariantized equations.

For comparison, we also considered the bosonic spin--2 case. There, the
simplest gravitational covariantization already fails on generic curved
backgrounds because the divergence of the wave equation produces both
\((Riemann)\cdot\nabla\Phi\) and \((\nabla Ricci)\cdot\Phi\) terms. On Einstein
spaces, the derivative-of-Ricci terms disappear, but an irreducible
Weyl-coupled derivative obstruction remains. Unlike the spin-$\tfrac32$
problem, this obstruction is not removed by a mass--charge tuning. Within the
local algebraic curvature deformations studied here, it is canceled uniquely by
a non-minimal Riemann coupling. The comparison is useful because it highlights a
genuine structural difference: the fermionic first-order system and the bosonic
second-order system solve their consistency problems in different ways.

Several directions remain open. First, it would be useful to perform a full
characteristic analysis of the tuned operator on Einstein--Maxwell backgrounds,
to connect the present lower-spin-chain derivation more directly to
hyperbolicity and causality. Second, one would like a systematic classification
of local first-order deformations beyond the reduced \(\beta_\kappa\)-family, in
order to understand whether any larger class can evade the rigidity of the
chiral scalar obstruction without spoiling the Einstein--Maxwell stress-tensor
matching.

The overall picture is therefore simple. In the decoupled regime, a charged
massive spin-$\tfrac32$ field admits a consistent \(g=2\) Fierz--Pauli
description with independent mass and charge. Once gravity is dynamical, the
covariant lower-spin chain forces a different organization. The gravitational
background does not merely deform the flat-space system: it probes a channel
that the maximally asymmetric organization cannot match. Within the local
two-derivative framework studied here, the supergravity endpoint is therefore
not just another realization of the same particle, but the unique organization
selected by gravitational backreaction. This also suggests that other avenues,
in particular string-theoretic completions involving Regge-trajectory states,
deserve further investigation as possible mechanisms for restoring consistency
beyond the minimal local two-derivative framework.

\appendix

\section{Conventions}
\label{app:conventions}

In this appendix we collect the conventions used throughout the paper.

\subsection*{Space-time conventions}

We work in four-dimensional Lorentzian signature
\begin{equation}
g_{mn}=\mathrm{diag}(+,-,-,-).
\end{equation}
Space-time indices are denoted by \(m,n,p,q,\ldots\).  Covariant derivatives are torsionless and
metric-compatible,
\begin{equation}
\nabla_m g_{np}=0.
\end{equation}
The gauge-covariant derivative acting on a field of electric charge \(e\) is
\begin{equation}
D_m=\nabla_m+i e A_m .
\end{equation}
The field strength is
\begin{equation}
F_{mn}=\partial_m A_n-\partial_n A_m.
\end{equation}

The commutator of gauge-covariant derivatives on a vector--spinor is
\begin{equation}
[D_m,D_n]\chi_p
=
-\,R_{mnp}{}^{q}\chi_q
+\frac14\,R_{mnrs}\,\sigma^{rs}\chi_p
+i e F_{mn}\chi_p,
\end{equation}
and similarly for dotted spinors with \(\bar\sigma^{rs}\) in place of \(\sigma^{rs}\).

The Ricci tensor and Einstein tensor are defined by
\begin{equation}
R_{mn}=R_{mpn}{}^{p},
\qquad
G_{mn}=R_{mn}-\frac12 g_{mn}R.
\end{equation}

\subsection*{Levi--Civita tensor and Hodge dual}

We take
\begin{equation}
\eps^{0123}=+1,
\qquad
\eps_{0123}=-1.
\end{equation}
The dual of the field strength is
\begin{equation}
\widetilde F^{mn}\equiv \frac12\,\eps^{mnrs}F_{rs},
\qquad
\widetilde{\widetilde F}_{mn}=-F_{mn}.
\end{equation}
We also use the self-dual and anti-self-dual combinations
\begin{equation}
(F_\pm)^{mn}\equiv F^{mn}\pm i\,\widetilde F^{mn}.
\end{equation}

The Maxwell stress tensor is
\begin{equation}
T^{(F)}_{mn}
=
F_{mr}F_n{}^{r}
-\frac14 g_{mn}\,F_{rs}F^{rs}.
\label{eq:app_maxwell_stress}
\end{equation}

\subsection*{Two-component Weyl spinors}

We use two-component Weyl notation. Undotted indices are
\(\alpha,\beta,\ldots\), dotted indices are
\(\dot\alpha,\dot\beta,\ldots\).  The Pauli matrices are
\begin{equation}
\sigma^m_{\alpha\dot\alpha}=(\mathbf 1,\vec\sigma),
\qquad
\bar\sigma^{m\,\dot\alpha\alpha}=(\mathbf 1,-\vec\sigma).
\end{equation}
They satisfy
\begin{equation}
\sigma^m_{\alpha\dot\alpha}\bar\sigma^{n\,\dot\alpha\beta}
+
\sigma^n_{\alpha\dot\alpha}\bar\sigma^{m\,\dot\alpha\beta}
=
2g^{mn}\,\delta_\alpha{}^\beta,
\end{equation}
\begin{equation}
\bar\sigma^{m\,\dot\alpha\alpha}\sigma^n_{\alpha\dot\beta}
+
\bar\sigma^{n\,\dot\alpha\alpha}\sigma^m_{\alpha\dot\beta}
=
2g^{mn}\,\delta^{\dot\alpha}{}_{\dot\beta}.
\end{equation}

The antisymmetric Lorentz generators are
\begin{equation}
\sigma^{mn}\equiv \frac14\bigl(\sigma^m\bar\sigma^n-\sigma^n\bar\sigma^m\bigr),
\qquad
\bar\sigma^{mn}\equiv \frac14\bigl(\bar\sigma^m\sigma^n-\bar\sigma^n\sigma^m\bigr).
\end{equation}
Useful identities are
\begin{equation}
\sigma^m\bar\sigma^n=g^{mn}\,\mathbf 1+2\sigma^{mn},
\qquad
\bar\sigma^m\sigma^n=g^{mn}\,\mathbf 1+2\bar\sigma^{mn},
\label{eq:app_sigmabar_sigma_two}
\end{equation}
and the triple-\(\sigma\) identities
\begin{equation}
\sigma^r\bar\sigma^s\sigma^n
=
g^{rs}\sigma^n-g^{rn}\sigma^s+g^{sn}\sigma^r
-i\,\eps^{rsn}{}_{m}\sigma^m,
\label{eq:app_triple_sigma_conv}
\end{equation}
\begin{equation}
\bar\sigma^r\sigma^s\bar\sigma^n
=
g^{rs}\bar\sigma^n-g^{rn}\bar\sigma^s+g^{sn}\bar\sigma^r
+i\,\eps^{rsn}{}_{m}\bar\sigma^m.
\label{eq:app_triple_barsigma_conv}
\end{equation}

Spinor indices are raised and lowered with
\(\eps_{\alpha\beta}\), \(\eps^{\alpha\beta}\),
\(\eps_{\dot\alpha\dot\beta}\), \(\eps^{\dot\alpha\dot\beta}\), with
\begin{equation}
\eps^{12}=+1,
\qquad
\eps_{12}=-1.
\end{equation}

\subsection*{Vector--spinor decomposition}

The four-component vector--spinor is decomposed as
\begin{equation}
\Psi_m=
\begin{pmatrix}
\chi_{m\alpha}\\[1mm]
\bar\lambda_m{}^{\dot\alpha}
\end{pmatrix}.
\end{equation}
The reduced spin-\(\tfrac32\) primary constraints are
\begin{equation}
\bar\sigma^m\chi_m=0,
\qquad
\sigma^m\bar\lambda_m=0.
\label{eq:app_primary_constraints_conv}
\end{equation}

The reduced secondary quantities used in the main text are
\begin{equation}
\fS^{(L)}_\alpha \equiv D^m\chi_{m\alpha},
\qquad
\fS^{(R)}_{\dot\alpha}
\equiv
D^m\bar\lambda_{m\dot\alpha}
+\frac{e}{M}(1-\bk)\,\bar\sigma^m_{\dot\alpha\alpha}F_{mn}\chi^{n\alpha}.
\end{equation}

\subsection*{Four-component notation}

When needed, we use the chiral representation for the Dirac matrices,
\begin{equation}
\gamma^m=
\begin{pmatrix}
0 & \sigma^m_{\alpha\dot\alpha}\\
\bar\sigma^{m\,\dot\alpha\alpha} & 0
\end{pmatrix},
\qquad
\gamma_5=
\begin{pmatrix}
-\mathbf 1 & 0\\
0 & \mathbf 1
\end{pmatrix}.
\end{equation}
The antisymmetrized products are
\begin{equation}
\gamma^{mn}\equiv \frac12[\gamma^m,\gamma^n],
\qquad
\gamma^{mnp}\equiv \gamma^{[m}\gamma^n\gamma^{p]}.
\end{equation}

\subsection*{Einstein--Maxwell background equations}

When specializing to Einstein--Maxwell backgrounds, we impose
\begin{equation}
\nabla_m F^{mn}=0,
\qquad
\nabla_{[m}F_{np]}=0,
\qquad
G_{mn}+\Lambda g_{mn}=\kappa^2 T^{(F)}_{mn}.
\label{eq:app_EM_background_eqs}
\end{equation}
Since \(T^{(F)}{}^m{}_m=0\) in four dimensions, this implies
\begin{equation}
R=4\Lambda,
\qquad
R_{mn}=\Lambda g_{mn}+\kappa^2 T^{(F)}_{mn}.
\end{equation}

\subsection*{Constant-field algebraic sector}

In several places we isolate the purely algebraic constant-field sector. By this we mean that
\begin{equation}
\nabla_p F_{mn}=0
\end{equation}
is imposed inside the specific algebraic step under consideration, so that terms involving
\(\nabla F\) are omitted there and treated separately in the later analysis of non-constant
electromagnetic backgrounds.

\section{Spinor identities used in the reduced analysis}
\label{app:spinor_identities}

In this appendix we collect the two-component identities used in the reduced
constraint analysis and prove them explicitly. Throughout we use the conventions of
App.~\ref{app:conventions}.

\subsection*{Basic \texorpdfstring{$\sigma$}{sigma}-matrix identities}

The Clifford relations are
\begin{equation}
\sigma^m\bar\sigma^n+\sigma^n\bar\sigma^m=2g^{mn}\,\mathbf 1,
\qquad
\bar\sigma^m\sigma^n+\bar\sigma^n\sigma^m=2g^{mn}\,\mathbf 1.
\label{eq:app_spinor_clifford}
\end{equation}
From these one obtains
\begin{equation}
\sigma^m\bar\sigma^n=g^{mn}\,\mathbf 1+2\sigma^{mn},
\qquad
\bar\sigma^m\sigma^n=g^{mn}\,\mathbf 1+2\bar\sigma^{mn},
\label{eq:app_spinor_sigmamn}
\end{equation}
with
\begin{equation}
\sigma^{mn}\equiv \frac14\bigl(\sigma^m\bar\sigma^n-\sigma^n\bar\sigma^m\bigr),
\qquad
\bar\sigma^{mn}\equiv \frac14\bigl(\bar\sigma^m\sigma^n-\bar\sigma^n\sigma^m\bigr).
\end{equation}

A second set of identities needed repeatedly is the triple-\(\sigma\) algebra:
\begin{equation}
\sigma^r\bar\sigma^s\sigma^n
=
g^{rs}\sigma^n-g^{rn}\sigma^s+g^{sn}\sigma^r
-i\,\eps^{rsn}{}_{m}\sigma^m,
\label{eq:app_spinor_triple_sigma}
\end{equation}
\begin{equation}
\bar\sigma^r\sigma^s\bar\sigma^n
=
g^{rs}\bar\sigma^n-g^{rn}\bar\sigma^s+g^{sn}\bar\sigma^r
+i\,\eps^{rsn}{}_{m}\bar\sigma^m.
\label{eq:app_spinor_triple_barsigma}
\end{equation}
These follow directly from \eqref{eq:app_spinor_clifford} by inserting
\(\sigma^r\bar\sigma^s=g^{rs}+2\sigma^{rs}\) and using the explicit form of the Lorentz
generators.

\subsection*{Chiral factorization identities}

The self-dual and anti-self-dual field strengths are defined by
\begin{equation}
(F_\pm)^{mn}\equiv F^{mn}\pm i\,\widetilde F^{mn},
\qquad
\widetilde F^{mn}\equiv \frac12\,\eps^{mnrs}F_{rs}.
\end{equation}
The key identities used in the reduced divergence analysis are the chiral factorizations
\begin{equation}
\sigma_m(F_+)^{mn}=F_{rs}\,\sigma^{rs}\sigma^n,
\label{eq:app_sigma_Fplus_factor}
\end{equation}
\begin{equation}
\bar\sigma_m(F_-)^{mn}=F_{rs}\,\bar\sigma^{rs}\bar\sigma^n.
\label{eq:app_barsigma_Fminus_factor}
\end{equation}

\paragraph{Proof of \eqref{eq:app_sigma_Fplus_factor}.}
Contract \eqref{eq:app_spinor_triple_sigma} with the antisymmetric tensor \(F_{rs}\):
\begin{equation}
F_{rs}\sigma^r\bar\sigma^s\sigma^n
=
F_{rs}\Bigl(
g^{rs}\sigma^n-g^{rn}\sigma^s+g^{sn}\sigma^r
-i\,\eps^{rsn}{}_{m}\sigma^m
\Bigr).
\end{equation}
The first term vanishes because \(F_{rs}\) is antisymmetric. The two metric contractions give
\begin{equation}
-\,F_{rs}g^{rn}\sigma^s+F_{rs}g^{sn}\sigma^r
=
2F^{mn}\sigma_m.
\end{equation}
For the Levi--Civita term we use
\begin{equation}
\eps^{rsn}{}_{m}F_{rs}=2\,\widetilde F^{n}{}_{m},
\end{equation}
hence
\begin{equation}
-\,i\,\eps^{rsn}{}_{m}F_{rs}\sigma^m
=
2i\,\widetilde F^{mn}\sigma_m.
\end{equation}
Combining the pieces,
\begin{equation}
F_{rs}\sigma^r\bar\sigma^s\sigma^n
=
2\bigl(F^{mn}+i\widetilde F^{mn}\bigr)\sigma_m
=
2\sigma_m(F_+)^{mn}.
\end{equation}
On the other hand, antisymmetry of \(F_{rs}\) implies
\begin{equation}
F_{rs}\sigma^r\bar\sigma^s
=
2F_{rs}\sigma^{rs}.
\end{equation}
This proves \eqref{eq:app_sigma_Fplus_factor}. \qed

\paragraph{Proof of \eqref{eq:app_barsigma_Fminus_factor}.}
The proof is identical, starting from \eqref{eq:app_spinor_triple_barsigma}. The sign of the
Levi--Civita term is reversed, which produces \(F_-^{mn}=F^{mn}-i\widetilde F^{mn}\). \qed

\subsection*{Primary-surface identities}

The factorizations above immediately imply the chiral trace identities used to derive the
secondary constraints.

If the primary constraint
\begin{equation}
\sigma^n\bar\lambda_n=0
\end{equation}
holds, then
\begin{equation}
\sigma_m(F_+)^{mn}\bar\lambda_n
=
F_{rs}\sigma^{rs}(\sigma^n\bar\lambda_n)
=0.
\label{eq:app_sigma_Fplus_primary_zero}
\end{equation}
Similarly, if
\begin{equation}
\bar\sigma^n\chi_n=0,
\end{equation}
then
\begin{equation}
\bar\sigma_m(F_-)^{mn}\chi_n
=
F_{rs}\bar\sigma^{rs}(\bar\sigma^n\chi_n)
=0.
\label{eq:app_barsigma_Fminus_primary_zero}
\end{equation}

These are precisely the identities used in the derivation of the secondary divergences.

\subsection*{Projected derivative identity}

The reduced closure analysis requires an identity for the contraction
\((F_+)^{mn}D_m\bar\lambda_n\). This is obtained by differentiating the chiral primary identity.

Starting from \eqref{eq:app_sigma_Fplus_primary_zero}, and restricting to the constant-field
algebraic sector \(\nabla_pF_{mn}=0\), one finds
\begin{equation}
\sigma_m(F_+)^{mn}D_p\bar\lambda_n=0.
\label{eq:app_diff_chiral_primary}
\end{equation}
Contracting with \(\bar\sigma^p\) and using
\begin{equation}
\bar\sigma^p\sigma_m=g^p{}_m\,\mathbf 1+2\,\bar\sigma^p{}_m,
\qquad
\bar\sigma^p{}_m\equiv \frac14\bigl(\bar\sigma^p\sigma_m-\bar\sigma_m\sigma^p\bigr),
\label{eq:app_projected_sigma_identity}
\end{equation}
gives
\begin{equation}
(F_+)^{pn}D_p\bar\lambda_n
+
2\,\bar\sigma^p{}_m(F_+)^{mn}D_p\bar\lambda_n
=0.
\label{eq:app_projected_derivative_identity}
\end{equation}
This is the identity used in the main text to isolate the Pauli-divergence contraction from the
reduced equation of motion.

The conjugate identity is obtained in the same way from
\eqref{eq:app_barsigma_Fminus_primary_zero}:
\begin{equation}
(F_-)^{pn}D_p\chi_n
+
2\,\sigma^p{}_m(F_-)^{mn}D_p\chi_n
=0,
\label{eq:app_projected_derivative_identity_conj}
\end{equation}
with
\begin{equation}
\sigma^p{}_m\equiv \frac14\bigl(\sigma^p\bar\sigma_m-\sigma_m\bar\sigma^p\bigr).
\end{equation}

\subsection*{Mixed quadratic identity}

The genuinely new algebraic tensor channel comes from the mixed product \(F_+F_-\). The relevant
identity is
\begin{equation}
(F_+)^{mn}(F_-)_{nq}=2\,T^{(F)m}{}_{q},
\label{eq:app_mixed_quadratic_identity}
\end{equation}
where the Maxwell stress tensor is
\begin{equation}
T^{(F)}_{mq}
=
F_{mr}F_q{}^{r}
-\frac14 g_{mq}\,F_{rs}F^{rs}.
\end{equation}

\paragraph{Proof.}
Expand the product:
\begin{equation}
(F_+)^{mn}(F_-)_{nq}
=
\bigl(F^{mn}+i\widetilde F^{mn}\bigr)
\bigl(F_{nq}-i\widetilde F_{nq}\bigr).
\end{equation}
This gives
\begin{equation}
F^{mn}F_{nq}
+\widetilde F^{mn}\widetilde F_{nq}
+i\bigl(\widetilde F^{mn}F_{nq}-F^{mn}\widetilde F_{nq}\bigr).
\end{equation}
The imaginary part vanishes identically. For the real part one uses the standard four-dimensional
identity
\begin{equation}
\widetilde F^{mn}\widetilde F_{nq}
=
F^{mn}F_{nq}
-\frac12\,\delta^m{}_q\,F_{rs}F^{rs}.
\end{equation}
Therefore
\begin{equation}
(F_+)^{mn}(F_-)_{nq}
=
2F^{mn}F_{nq}
-\frac12\,\delta^m{}_q\,F_{rs}F^{rs}
=
2\,T^{(F)m}{}_{q},
\end{equation}
which proves \eqref{eq:app_mixed_quadratic_identity}. \qed

As an immediate consequence,
\begin{equation}
(F_+)^{mn}\bar\sigma_m(F_-)_{nq}\chi^q
=
2\,\bar\sigma_m T^{(F)m}{}_{q}\chi^q
=
2\,\bar\sigma^mT^{(F)}_{mn}\chi^n,
\label{eq:app_mixed_quadratic_consequence_left}
\end{equation}
and similarly
\begin{equation}
(F_-)^{mn}\sigma_m(F_+)_{nq}\bar\lambda^q
=
2\,\sigma^mT^{(F)}_{mn}\bar\lambda^n.
\label{eq:app_mixed_quadratic_consequence_right}
\end{equation}

Equations \eqref{eq:app_sigma_Fplus_factor}--\eqref{eq:app_mixed_quadratic_consequence_right}
are the identities used in Secs.~\ref{sec:beta_kappa_reduced_system} and
\ref{sec:beta_kappa_closure_core}.

\section{Two-component identities used in the reduced closure analysis}
\label{app:2c_identities}

In this appendix we collect the two-component identities used in the reduced
constraint analysis and prove them explicitly.  Throughout we work in four
dimensions with mostly-minus signature and define
\begin{equation}
\widetilde F^{mn}\equiv \frac12\,\eps^{mnrs}F_{rs},
\qquad
(F_\pm)^{mn}\equiv F^{mn}\pm i\,\widetilde F^{mn},
\end{equation}
together with
\begin{equation}
\sigma^{mn}\equiv \frac14\bigl(\sigma^m\bar\sigma^n-\sigma^n\bar\sigma^m\bigr),
\qquad
\bar\sigma^{mn}\equiv \frac14\bigl(\bar\sigma^m\sigma^n-\bar\sigma^n\sigma^m\bigr).
\end{equation}

\subsection*{Triple-\texorpdfstring{$\sigma$}{sigma} identities}

We begin from the standard Clifford identities
\begin{equation}
\sigma^m\bar\sigma^n+\sigma^n\bar\sigma^m = 2g^{mn}\,\mathbf 1,
\qquad
\bar\sigma^m\sigma^n+\bar\sigma^n\sigma^m = 2g^{mn}\,\mathbf 1.
\label{eq:app_clifford_basic}
\end{equation}
From these one derives the triple-\(\sigma\) formulas
\begin{equation}
\sigma^r\bar\sigma^s\sigma^n
=
g^{rs}\sigma^n-g^{rn}\sigma^s+g^{sn}\sigma^r
-i\,\eps^{rsn}{}_{m}\sigma^m,
\label{eq:app_triple_sigma}
\end{equation}
\begin{equation}
\bar\sigma^r\sigma^s\bar\sigma^n
=
g^{rs}\bar\sigma^n-g^{rn}\bar\sigma^s+g^{sn}\bar\sigma^r
+i\,\eps^{rsn}{}_{m}\bar\sigma^m.
\label{eq:app_triple_barsigma}
\end{equation}
These are the only ingredients needed below.

\subsection*{Chiral factorization of \texorpdfstring{$F_\pm$}{Fpm}}

The reduced secondary constraints and the projected-divergence manipulations rely
on the fact that \(F_+\) and \(F_-\) factor through the chiral traces.

\paragraph{Identity 1.}
\begin{equation}
\sigma_m(F_+)^{mn}=F_{rs}\,\sigma^{rs}\sigma^n .
\label{eq:app_sigmaFplus_factorized}
\end{equation}

\paragraph{Proof.}
Contract \eqref{eq:app_triple_sigma} with the antisymmetric tensor \(F_{rs}\):
\begin{equation}
F_{rs}\sigma^r\bar\sigma^s\sigma^n
=
F_{rs}\Bigl(
g^{rs}\sigma^n-g^{rn}\sigma^s+g^{sn}\sigma^r
-i\,\eps^{rsn}{}_{m}\sigma^m
\Bigr).
\end{equation}
The term proportional to \(g^{rs}\) vanishes because \(F_{rs}\) is antisymmetric.
The two metric contractions give
\begin{equation}
-\,F_{rs}g^{rn}\sigma^s + F_{rs}g^{sn}\sigma^r
=
2\,F^{mn}\sigma_m .
\end{equation}
For the Levi--Civita term we use
\begin{equation}
\eps^{rsn}{}_{m}F_{rs}=2\,\widetilde F^{n}{}_{m},
\end{equation}
and therefore
\begin{equation}
-\,i\,\eps^{rsn}{}_{m}F_{rs}\sigma^m
=
2i\,\widetilde F^{mn}\sigma_m .
\end{equation}
Combining the pieces,
\begin{equation}
F_{rs}\sigma^r\bar\sigma^s\sigma^n
=
2\bigl(F^{mn}+i\widetilde F^{mn}\bigr)\sigma_m
=
2\,\sigma_m(F_+)^{mn}.
\end{equation}
Using antisymmetry of \(F_{rs}\),
\begin{equation}
F_{rs}\sigma^r\bar\sigma^s
=
2F_{rs}\sigma^{rs},
\end{equation}
which proves \eqref{eq:app_sigmaFplus_factorized}. \qed

\medskip

\paragraph{Identity 2.}
\begin{equation}
\bar\sigma_m(F_-)^{mn}=F_{rs}\,\bar\sigma^{rs}\bar\sigma^n .
\label{eq:app_barsigmaFminus_factorized}
\end{equation}

\paragraph{Proof.}
The proof is identical, starting instead from
\eqref{eq:app_triple_barsigma}. The sign flip in the Levi--Civita term produces
\(F_-^{mn}=F^{mn}-i\widetilde F^{mn}\). \qed

\subsection*{Vanishing of the chiral traces on the primary surface}

The previous factorization identities immediately imply the chiral trace relations
used throughout the reduced analysis.

\paragraph{Identity 3.}
If
\begin{equation}
\sigma^n\bar\lambda_n=0,
\label{eq:app_primary_right}
\end{equation}
then
\begin{equation}
\sigma_m(F_+)^{mn}\bar\lambda_n=0.
\label{eq:app_sigmaFplus_trace_zero}
\end{equation}

\paragraph{Proof.}
Using \eqref{eq:app_sigmaFplus_factorized},
\begin{equation}
\sigma_m(F_+)^{mn}\bar\lambda_n
=
F_{rs}\sigma^{rs}\sigma^n\bar\lambda_n.
\end{equation}
The right-hand side vanishes identically by the primary constraint
\eqref{eq:app_primary_right}. \qed

\medskip

\paragraph{Identity 4.}
If
\begin{equation}
\bar\sigma^n\chi_n=0,
\label{eq:app_primary_left}
\end{equation}
then
\begin{equation}
\bar\sigma_m(F_-)^{mn}\chi_n=0.
\label{eq:app_barsigmaFminus_trace_zero}
\end{equation}

\paragraph{Proof.}
Using \eqref{eq:app_barsigmaFminus_factorized},
\begin{equation}
\bar\sigma_m(F_-)^{mn}\chi_n
=
F_{rs}\bar\sigma^{rs}\bar\sigma^n\chi_n,
\end{equation}
which vanishes by \eqref{eq:app_primary_left}. \qed

These are the identities used in Sec.~\ref{sec:beta_kappa_reduced_system} to derive the
secondary divergences.

\subsection*{Projected identity for \texorpdfstring{$(F_+)^{mn}D_m\bar\lambda_n$}{Fplus D lambdabar}}

In Sec.~\ref{sec:beta_kappa_closure_core} we needed a more refined identity to isolate the
contraction \((F_+)^{mn}D_m\bar\lambda_n\) appearing in the Pauli divergence.

On the primary surface,
\begin{equation}
\sigma_m(F_+)^{mn}\bar\lambda_n=0.
\end{equation}
In the constant-field algebraic sector, where \(\nabla_pF_{mn}=0\), differentiating gives
\begin{equation}
\sigma_m(F_+)^{mn}D_p\bar\lambda_n=0.
\label{eq:app_diff_primary_projected}
\end{equation}
Contract with \(\bar\sigma^p\) and use
\begin{equation}
\bar\sigma^{pm}\equiv \frac14\bigl(\bar\sigma^p\sigma^m-\bar\sigma^m\sigma^p\bigr),
\qquad
\bar\sigma^p\sigma_m=g^p{}_m\,\mathbf 1+2\bar\sigma^p{}_m .
\label{eq:app_bar_sigma_pm_def}
\end{equation}
Then \eqref{eq:app_diff_primary_projected} becomes
\begin{equation}
(F_+)^{pn}D_p\bar\lambda_n
+
2\,\bar\sigma^p{}_m(F_+)^{mn}D_p\bar\lambda_n
=0.
\label{eq:app_projected_FplusDlambda_identity}
\end{equation}
This is the projection formula used to replace the Pauli-divergence contraction by a
\(\bar\sigma\)-projected derivative, which can then be matched to the reduced equation of motion.

The conjugate identity is obtained similarly from
\(\bar\sigma_m(F_-)^{mn}\chi_n=0\):
\begin{equation}
(F_-)^{pn}D_p\chi_n
+
2\,\sigma^p{}_m(F_-)^{mn}D_p\chi_n
=0,
\label{eq:app_projected_FminusDchi_identity}
\end{equation}
again in the constant-field algebraic sector.

\subsection*{Mixed quadratic identity and the Maxwell stress tensor}

The new algebraic obstruction arises from the mixed quadratic product
\(F_+F_-\).  The required identity is
\begin{equation}
(F_+)^{mn}(F_-)_{nq}=2\,T^{(F)m}{}_{q},
\label{eq:app_mixed_quadratic_identity}
\end{equation}
where
\begin{equation}
T^{(F)}_{mq}
=
F_{mr}F_q{}^{r}
-\frac14 g_{mq}\,F_{rs}F^{rs}.
\end{equation}

\paragraph{Proof.}
Expand
\begin{equation}
(F_+)^{mn}(F_-)_{nq}
=
\bigl(F^{mn}+i\widetilde F^{mn}\bigr)
\bigl(F_{nq}-i\widetilde F_{nq}\bigr).
\end{equation}
This gives
\begin{equation}
F^{mn}F_{nq}
+\widetilde F^{mn}\widetilde F_{nq}
+i\bigl(\widetilde F^{mn}F_{nq}-F^{mn}\widetilde F_{nq}\bigr).
\end{equation}
The mixed imaginary term vanishes identically. For the real part, use the standard four-dimensional identity
\begin{equation}
\widetilde F^{mn}\widetilde F_{nq}
=
F^{mn}F_{nq}
-\frac12\,\delta^m{}_q\,F_{rs}F^{rs}.
\end{equation}
Therefore
\begin{equation}
(F_+)^{mn}(F_-)_{nq}
=
2F^{mn}F_{nq}
-\frac12\,\delta^m{}_q\,F_{rs}F^{rs}
=
2\,T^{(F)m}{}_{q}.
\end{equation}
This proves \eqref{eq:app_mixed_quadratic_identity}. \qed

As a consequence,
\begin{equation}
(F_+)^{mn}\bar\sigma_m(F_-)_{nq}\chi^q
=
2\,\bar\sigma_m\,T^{(F)m}{}_{q}\chi^q
=
2\,\bar\sigma^mT^{(F)}_{mn}\chi^n,
\label{eq:app_sigma_mixed_quadratic_consequence}
\end{equation}
and similarly
\begin{equation}
(F_-)^{mn}\sigma_m(F_+)_{nq}\bar\lambda^q
=
2\,\sigma^mT^{(F)}_{mn}\bar\lambda^n .
\label{eq:app_barsigma_mixed_quadratic_consequence}
\end{equation}

Equations \eqref{eq:app_projected_FplusDlambda_identity}--
\eqref{eq:app_barsigma_mixed_quadratic_consequence} are the identities used in
Sec.~\ref{sec:beta_kappa_closure_core} to extract the Einstein--Maxwell dressed
secondary obstruction.

\end{document}